\documentclass[12pt, onecolumn,draftcls]{IEEEtran}
\usepackage{amsmath}
\usepackage[mathscr]{eucal}
\usepackage{amsfonts}
\usepackage{amssymb}
\usepackage{psfrag}
\usepackage{cite}
\usepackage{color}
\usepackage{graphicx}

\input{TNDs_IEEEtran_v16_mods}
\newcommand{\paperstatus}{}

\newtheorem{theorem}{Theorem}
\newtheorem{lemma}{Lemma}

\newtheorem{assumption}{Assumption}
\newtheorem{problem}{Problem}

\newtheorem{proposition}{Proposition}

\newtheorem{property}{Property}

\newcommand{\smatlabaxislabel}[1]{\fontsize{12}{\f@baselineskip}%
\textsf{#1}}
\newcommand{\matlabaxislabel}[1]{\fontsize{14.4}{\f@baselineskip}%
\textsf{#1}}
\newcommand{\mmatlabaxislabel}[1]{\fontsize{17.28}{\f@baselineskip}%
\textsf{#1}}
\newcommand{\bmatlabaxislabel}[1]{\fontsize{20.74}{\f@baselineskip}%
\textsf{#1}}
\newcommand{\bbmatlabaxislabel}[1]{\fontsize{24.88}{\f@baselineskip}%
\textsf{#1}} \makeatother

\allowdisplaybreaks

\centerfigcaptionstrue

\newcommand{\half}%
{\raisebox{2.5pt}{\scriptsize 1}{\small
/}\raisebox{-1pt}{\scriptsize 2}}

\newcommand{\beq}{\begin{equation}}
\newcommand{\eeq}{\end{equation}}

\title{Space Codes for MIMO Optical Wireless Communications: Error Performance Criterion and Code Construction}
\author{Yan-Yu Zhang\\Zhengzhou Information Science and Technology Institute}
\author{Yan-Yu Zhang, Hong-Yi Yu, Jian-Kang Zhang, Senior Member, IEEE, Yi-Jun Zhu, Jin-Long Wang and Tao Wang
\thanks{Partial results of this paper has been presented in ISIT 2015, Hong Kong. This work was supported in part by   Key Laboratory of Universal Wireless  Communications (BUPT), Ministry of Education of P. R. China under Grant No. KFKT-2012103, in part by NNSF of China (No. 61271253), and  in part by NHTRDP (863 Program) of China (Grant No.2013AA013603).}
\thanks{
Yan-Yu Zhang, Hong-Yi Yu, Yi-Jun Zhu, Jin-Long Wang and Tao Wang
are with Department of Communication Engineering, Zhengzhou Information Science and Technology
Institute, Zhengzhou, Henan Province (450000), China. Emails:
yyzhang.xinda@gmail.com, maxyucn@sohu.com, yijunzhu1976@gmail.com, wjl543@sina.com and yjswangtao@163.com.}
\thanks{Jian-Kang Zhang is with the Department of Electrical and Computer Engineering,
McMaster University, 1280 Main Street West, L8S 4K1, Hamilton,
Ontario, Canada. Email: jkzhang@mail.ece.mcmaster.ca.
}}

\begin{document}
\maketitle

\begin{abstract}
In this paper, we consider a multiple-input-multiple-output optical wireless communication (MIMO-OWC) system in the presence of log-normal fading. In this scenario, a general  criterion for the design of full-diversity  space code (FDSC) with the maximum likelihood (ML) detector is developed. This criterion  reveals that in a high signal-to-noise ratio (SNR) regime, MIMO-OWC  offers  both large-scale diversity gain, governing the exponential decaying of the error curve, and small-scale diversity gain, producing traditional power-law decaying. Particularly for a two by two MIMO-OWC system with  unipolar pulse amplitude modulation (PAM), a closed-form solution to the design problem of a linear FDSC optimizing both diversity gains is attained by taking advantage of the available properties on the successive terms of Farey sequences in number theory as well as by developing new properties on the disjoint intervals formed by the Farey sequence terms  to attack the continuous and discrete variables mixed max-min design problem. In fact, this specific design not only proves that a repetition code (RC) is the optimal linear FDSC optimizing both the diversity gains, but also uncovers a significant difference between MIMO radio frequency (RF) communications and  MIMO-OWC that space dimension alone is sufficient for a full large-scale diversity achievement. Computer simulations demonstrate that  FDSC substantially outperforms uncoded spatial multiplexing with the same total optical power and spectral efficiency,  and the latter  provides  only the small-scale diversity gain.
\end{abstract}
\section{Introduction}\label{sec:introduction}
In the past decade, the demand for capacity in cellular and wireless local area networks has grown in an explosive manner. This demand has triggered off an enormous expansion in radio frequency (RF) wireless communications. As an adjunct or alternative to RF communication, optical wireless communications (OWC), due to its potential for bandwidth-hungry applications,  has become a very important area of research~\cite{li2003optical,o2005optical,boucouvalas2005challenges,chan2006free,das2008requirements,Brien2008challenges,Lubin2009MIMO,Kumar2010Led-based,Elgala2011review,Borah2012review, Gancarz13,png2013mimo}. The importance of OWC lies in the advantages of low cost, high security, freedom from spectral
licensing issues etc. Furthermore, OWC links of practical interest involve satellites, deep-space probes, ground stations, unmanned aerial vehicles, high altitude platforms, aircraft, and other nomadic communication partners. Moreover, all these links can be used in both military and civilian contexts, or both indoor and outdoor scenarios in demand of high data rate. Therefore, OWC is considered to be the next frontier for net-centric connectivity for  bandwidth, spectrum and security issues.

 However, some challenges remain, especially in the mobile or atmospheric environments. For high data rate OWC systems over mobile or atmospheric channels, \textit{robustness} is a key consideration. In  mobile links, there will be inevitable impairments such as terminal-sway, aerosol scattering  and non-zero pointing errors~\cite{farid2007outage,borah2009pointing,Xuegui13,yang2014free}, etc. In addition, for  atmospheric environments, atmospheric effects, such as rain, snow, fog and temperature variation, will affect the  link performance. Therefore, in the design of OWC links, we need to consider these impairments-induced fading~\cite{roth2000review}. This fading of the received intensity signal can be described by the log-normal (LN) statistical model ~\cite{haas2002space,Liu2004itct,Filho2005el,navidpour2007itwc,Beaulieu2008itct,giggenbach2008fading}, which is considered in this paper. To combat fading, multi-input-multi-output (MIMO) OWC (MIMO-OWC) systems provide diverse replicas of transmitted symbols to the receiver by using multiple receiver apertures with sufficient separation between each so that the fading for each receiver is independent of others. Such diversity can also be achieved  by  introducing a design for the transmitted symbols distributed over transmitting apertures (space) and (or) symbol periods (time). Full diversity is achieved when  the decaying speed of the error curve for the coded MIMO-OWC system is maximized.

Unfortunately, unlike MIMO techniques for  radio frequency (MIMO-RF) communications with Rayleigh fading, there are \textit{two} significant challenges in MIMO-OWC communications. The \textit{first} is that there does not exist any available mathematical tool that could be directly applied to the analysis of the average pair-wise error probability (PEP) when LN is involved. In this scenario, it is indeed a challenge to extract a dominant term of the average PEP. Let alone say how to achieve a full diversity gain. Here, it should be mentioned that there are really  mathematical formulae  in literature for numerically and accurately computing the integral involving  LN~\cite{haas2002space,Liu2004itct,Filho2005el,navidpour2007itwc,Beaulieu2008itct}. However, it can not be used for the theoretic analysis on diversity. The \textit{second} is a \textit{nonnegative constraint} on the design of transmission for MIMO-OWC, which is a major difference between MIMO RF communications and MIMO-OWC. It is because of this constraint that the currently available well-developed MIMO techniques for RF communications can not be directly utilized for MIMO-OWC.  Despite the fact that the nonnegative constraint can be satisfied by properly adding some direct-current components (DC) into transmitter designs so that the existing advanced MIMO techniques~\cite{tarokh98} for RF communications such as orthogonal space-time block code (OSTBC)~\cite{geramita79,alamouti98,tarokh99,su01,ganesan01,tirkkonen02,liang03,liu-icassp04,cui07} could be used in MIMO-OWC, the power loss arising from DC incurs the fact that these modified OSTBCs~\cite{simon2005alamouti,wang2009mimo,tianhigh} in an LN fading optical channel have worse error performance than the RC~\cite{navidpour2007itwc,majid2008twc,bayaki2010space,abaza2014diversity}.

All the aforementioned factors greatly motivate us to develop a general criterion on the design of full-diversity transmission for MIMO-OWC. As an initial exploration, we consider the utilization of a space dimension alone,  and intend  to uncover some unique characteristics of MIMO-OWC. With this goal in mind, our main tasks in this paper are as follows.
\begin{enumerate}
  \item \textit{To establish a general criterion for the design of full-diversity space code (FDSC)}. To this end, our main idea here is that by fragmenting the integral region of the average PEP involving LN into two sub-domains adaptively with SNR, the dominant term will be extracted. With this, we will give a necessary and sufficient condition for a space code to assure full diversity.
  \item \textit{To attain an optimal analytical solution to a specific two by two linear FDSC design problem}. To do that, we formulate and simplify our optimization problem with continuous and discrete mixed design variables by taking advantage of available properties as well as by developing new properties on Farey sequences in number theory for our purpose. In fact, we will prove that  the RC is  the optimal linear FDSC for this specific scenario in the sense of the design criterion proposed in this paper.
\end{enumerate}

\section{Channel Model And Space Code}\label{sec:model}
In this section, we first briefly review the channel model which is considered in this paper. Then, we propose the space coding structure  and formulate the design problems to be solved.

\subsection{Channel Model with Space Code}\label{subsec:model}
Let us consider an $M\times N$ MIMO-OWC system having $M$ receiver apertures and $N$ transmitter apertures transmitting the symbol vector $\mathbf{s}$, $\{s_\ell\},\ell=1,~\cdots,~L$, which are randomly, independently and equally likely, selected from a given constellation. To facilitate the transmission of these $L$ symbols through the $N$ transmitters in the one time slot (channel use), each symbol $s_\ell$ is mapped by a space encoder $\mathbf{F}_{\ell}$ to an $N\times 1$  space code vector $\mathbf{F}_\ell\left(s_\ell\right)$ and then summed together, resulting in an $N\times 1$ space codeword given by $\mathbf{x}=\sum_{\ell=1}^L\mathbf{F}_\ell\left(s_\ell\right)$, where the $n$-th element of $\mathbf{x}$ represents the coded symbol to be transmitted from the $n$-th transmitter aperture. These coded symbols are then transmitted to the receivers through flat-fading path coefficients, which form the elements of the $M\times N$ channel matrix $\mathbf{H}$. The received space-only symbol, denoted by the $M\times 1$ vector $\mathbf{y}$, can be written as
 \begin{eqnarray}\label{eqn:system_model}
\mathbf{y}=\frac{1}{P_{op}}\mathbf{H}\mathbf{x}+\mathbf{n},
\end{eqnarray}
  where $P_{op}$ is the average optical power of $\mathbf{x}$ and,  the entries of  channel matrix $\mathbf{H}$ are independent and LN distributed, i.e., $h_{ij}=e^{z_{ij}}$, where $z_{ij}\sim\mathcal{N}\left(\mu_{ij},\sigma_{ij}^2\right), i=1,~\cdots,~M,j=1,~\cdots,~N$.
  The probability density function (PDF) of $h_{ij}$ is
\begin{eqnarray}
f_{H}\left(h_{ij}\right)=\frac{1}{\sqrt{2\pi}h_{ij}\sigma_{ij} }\exp\left(-\frac{\left(\ln h_{ij}-\mu_{ij}\right)^2}{2\sigma _{ij}^{2}}\right)
\end{eqnarray}
The PDF of $\mathbf{H}$ is $f_{\mathbf{H}}\left(\mathbf{H}\right)=\prod_{i=1}^{M}\prod_{j=1}^{N}f_{H}\left(h_{ij}\right)$.

  The signalling scheme of $\mathbf{s}$ is unipolar pulse amplitude modulation (PAM) to meet the unipolarity requirement of intensity modulation (IM), i.e., $\mathbf{s}\in\mathbb{R}_+^{L\times 1}$. As an example, the constellation of unipolar $2^p$-ary PAM is $\mathcal{B}_{2^p}=\{0,1,~\cdots,~2^p-1\}$, where $p$ is a positive integer. Then, the equivalent constellation of $\mathbf{s}$ is $\mathcal{S}=\{\mathbf{s}:s_i\in \mathcal{B}_{2^p},i=1,~\cdots,~L\}$, i.e., ${\mathcal S}={\mathcal B}_{2^p}^L$.

Furthermore, for noise vector $\mathbf{n}$, the two primary sources at the receiver front end are due to noise from the receiver electronics and shot noise from the received DC photocurrent induced by background  radiation~\cite{Karp1988,Barry1994Ifrd}. Although the signal intensity also results in shot noise, which is signal-dependent, the shot noise from hight-intensity background radiation dominates. Therefore, by the central limit theorem, this high-intensity shot noise for the lightwave-based OWC is closely approximated as additive, signal-independent, white, Gaussian noise (AWGN)~\cite{Barry1994Ifrd} with zero mean and variance $\sigma _{\mathbf{n}}^{2}$.

 By rewriting the channel matrix  as a vector and aligning the code-channel product to form a new channel vector, we have $\mathbf{Hx}=\left(\mathbf{I}_{M}\otimes\mathbf{x}^T\right)\textrm{vec}\left(\mathbf{H}\right)
$, where $\otimes$ denotes the Kronecker product operation and $\textrm{vec}\left(\mathbf{H}\right)=\left[h_{11},\ldots,h_{1N},\ldots,h_{M1},\ldots,h_{MN}\right]^T$.
For discussion convenience, we call $\mathbf{I}_{M}\otimes\mathbf{x}^T$ a codeword matrix. Then, the correlation matrix of the corresponding error coding matrix between $\mathbf{I}_{M}\otimes\mathbf{x}^T$ and $\mathbf{I}_{M}\otimes\tilde{\mathbf{x}}^T$ is given by
   \begin{eqnarray}\label{eqn:rank_one_equivalence}
           \left(\mathbf{I}_{M}\otimes\mathbf{x}^T-\mathbf{I}_{M}\otimes\tilde{\mathbf{x}}^T\right)^T\left(\mathbf{I}_{M}\otimes\mathbf{x}^T-\mathbf{I}_{M}\otimes\tilde{\mathbf{x}}^T\right)=
\mathbf{I}_{M}\otimes\left(\mathbf{e}\mathbf{e}^T\right)
   \end{eqnarray}
   where $\mathbf{e}$ is the ``distance'' between distinct codewords $\mathbf{x}$ and $\tilde{\mathbf{x}}$. All these non-zero $\mathbf{e}$ form an error set, denoted by $\mathcal{E}$.

   \subsection{Problem Statement}
To formally state our problem, we make the following assumptions throughout this paper.
\begin{enumerate}
\item \textit{Power constraint}. The average optical power is constrained, i.e., $E\left[\sum_{i=1}^{N}x_{i}\right]=P_{op}$. Although limits are placed on both the average and peak optical power transmitted, in the case of most practical modulated optical sources, it is the average optical power constraint that dominates~\cite{hranilovic2003optical}.
\item \textit{SNR  definition}. The optical SNR is defined by $\rho_{op}=\frac{1}{\sqrt{N\sigma_\mathbf{n}^2}}$, since the noise variance per dimension is assumed to be $\sigma_\mathbf{n}^2$. Thus, in expressions on error performance involved in the squared Euclidean distance, the term $\rho$, in fact, is equal to
     \begin{eqnarray}\label{eqn:electrical_snr}
       \rho=\rho_{op}^2=\frac{1}{N\sigma_\mathbf{n}^2}
     \end{eqnarray}
     with optical power being normalized by $\frac{1}{P_{op}}$. Unless stated otherwise, $\rho$ is referred to as  the squared optical SNR thereafter.
    \item\textit{Channel  state information}. Channel state information (CSI) at the receiver (CSIR) is known and  CSI at  the transmitter (CSIT) is unavailable.
\end{enumerate}

 Under the above  assumptions, our primary task in this paper is to establish a general criterion on the design of FDSC and  solve the following problem.
\begin{problem}\label{prob:design_problem} Design the space encoder $\mathbf{F}(\cdot)$ subject to the total optical power such that 1) $\forall \mathbf{s}\in \mathcal{S}, \mathbf{F}\left(\mathbf{s}\right) $ meets the unipolarity requirement of IM; 2) Full diversity is enabled for the ML receiver.~\hfill\QED
\end{problem}

Naturally, two questions come up immediately: 1) \textit{What is full diversity  referred to as for MIMO-OWC ?} 2) \textit{What is the design criterion ?} So far, both questions remain open and thus, motive us to analyse the error performance in the ensuing section.

\section{ Error Performance Analysis And Design Criterion for Space Code}\label{sec:performance_analysis}
In this section, our purpose is to  analyze the error performance of the ML detector and develop a design criterion for FDSC. To make our presentation more clear as well as our main idea more easily understandable, we begin with the symbol error probability (SEP) analysis on single-input-single-output OWC (SISO-OWC).

   \subsection{SEP of SISO-OWC with PAM }\label{subsec:siso_sep}

Recall that the PDF of a scalar channel $h$ is
$f_{H}\left(h\right)=\frac{1}{\sqrt{2\pi}h\sigma }\exp\left(-\frac{\left(\ln h\right)^2}{2\sigma ^{2}}\right)$. In this specific case, it is known~\cite{Proakis00} that the average SEP for $2^p$-ary PAM is given by
 \begin{eqnarray}\label{eqn:ser-siso}
  P_e\left(\rho\right) = c_1\int_{0}^{\infty}Q\left(\frac{\sqrt{c_2\rho }h}{P_{op}}\right)f_{H}\left(h\right) dh
\end{eqnarray}
where $\rho$ denotes the squared optical SNR defined by \eqref{eqn:electrical_snr}, $c_1=2-2^{1-p}$ and $c_2=\frac{3p}{2^{2p}-1}$. For  the sake of simplicity and without loss of generality, we assume $p=1$ which is the case of on-off keying (OOK).

\begin{lemma}\label{lemma:siso_sep}
For OOK, $P_e\left(\rho\right)$ is bounded by
 \begin{eqnarray}
  &&\frac{Q\left(\frac{1}{2\sigma}\right)}{\sqrt{2\pi}}\frac{1}{\ln\rho-\ln \frac{4\sigma^{2}}{P_{op}^2}}e^{-\frac{\left(\ln\rho -\ln\frac{P_{op}^2}{4\sigma^{2}}\right)^2}{8\sigma^2}}
  \le P_e\left(\rho\right)\nonumber\\
  &&\le \frac{1}{2}\left(\ln\rho\right)^{-1} e^{-\frac{\left(\ln\frac{\rho}{\ln^2\rho}-\ln\frac{ P^2_{op}}{\sigma^2}\right)^2}{8\sigma ^2}}
  +\mathcal{O}\left(e^{-\frac{\left(\ln\rho\right)^2}{8\sigma^2}}\right)
  \end{eqnarray}
~\hfill\QED
\end{lemma}

The proof of Lemma~\ref{lemma:siso_sep} is postponed into Appendix~\ref{app:siso_sep}.

In the proof of Lemma~\ref{lemma:siso_sep}, our selection of $\tau$ is reasonable in finding the dominant term adaptively with SNR.  Here, it should mentioned that the term ``adaptive'' is referred to the fact that the fragmentation of the integral region is done by a function of SNR. We also noticed that the splitting method is similar to those used in~\cite{wangtc03,yang2014free}. However, the significant difference between our proposed technique and those used in~\cite{wangtc03,yang2014free} is that our proposed integral-splitting method is to extract the dominant behavior by selecting $\tau$ adaptively with SNR. The SEP for SISO-OWC has been given and the adaptive fragmentation with SNR can help attack the integral involved in LN. In light of this, we extend this technique to the case of MIMO-OWC.
   \subsection{PEP of MIMO-OWC }\label{subsec:pep_mimo_ovlc}
This subsection aims at deriving the PEP of MIMO-OWC by means of the above-mentioned adaptive fragmentation with SNR and then, establishing a general design criterion for the space coded MIMO-OWC system.

 Given a channel realization $\mathbf{H}\in\mathbb{R}_{+}^{M\times N}$ and a transmitted signal vector $\mathbf{s}$, the probability
   of transmitting $\mathbf{s}$ and deciding in favor of $\hat{\mathbf s}$ with the ML receiver is given by~\cite{forney98}
\begin{eqnarray}\label{eqn:ml_detection_pep1}
P\left(\mathbf{s}\rightarrow\hat{\mathbf{s}}|\mathbf{H}\right)=Q\left(\frac{d\left(\mathbf{e}\right)}{2}\right) \end{eqnarray}
where $d^2\left(\mathbf{e}\right)=\frac{\rho}{NP_{op}^2}\sum_{i=1}^M\left(\mathbf{h}_i^T\mathbf{e}\right)^2$ with $\mathbf{h}_i=\left[h_{i1},~\cdots,~h_{iN}\right]^T,i=1,~\cdots,~M$. Averaging \eqref{eqn:ml_detection_pep1}  over $\mathbf{H}$ yields
\begin{eqnarray} \label{eqn:ml_detection_pep2}
P\left(\mathbf{s}\rightarrow\hat{\mathbf{s}}\right)
&=&\int P\left(\mathbf{s}\rightarrow\hat{\mathbf{s}}|\mathbf{H}\right)f_{\mathbf{H}}\left(\mathbf{H}\right)d\mathbf{H}.
\end{eqnarray}
For presentation clarity, we first give the definition of the dominant term.
 Suppose that there are two terms $T_1(\rho)$ and $T_2(\rho)$, each of which goes to zero when $\rho$ tends to infinity. We say that $T_1(\rho)$ is  a dominant term in the sum of $T_1(\rho)+T_2(\rho)$ if $\lim_{\rho\rightarrow\infty}T_2(\rho)/T_1(\rho)=0$.
To extract the dominant term of~\eqref{eqn:ml_detection_pep2},  we make an assumption for time being. Later on, we will prove that this condition is actually necessary and sufficient for a  full diversity achievement.

\begin{assumption} \label{assumpt:existence_of_rectangular}
Any $\mathbf{e}\in {\mathcal E}$ is unipolar without zero entry.~\hfill\QED
\end{assumption}
We are now in a position to formally state the first main result in this paper.
\begin{theorem}\label{theorem:mimo_pep}
Under Assumption~\ref{assumpt:existence_of_rectangular}, $P\left(\mathbf{s}\rightarrow\hat{\mathbf{s}}\right)$ is asymptotically bounded by
\begin{eqnarray}\label{eqn:pep_asymp}
  &&\underbrace{C_{L} \left(\ln\rho\right)^{-MN}e^{-\sum_{i=1}^{M}\sum_{j=1}^{N}\frac{\left(\ln\rho +\ln \left(P_{op}^2\Omega\right)-\ln\left(M\sum_{k=1}^Ne_k^2\right)\right)^2}{8\sigma_{ij}^2}}}_{P_{L}\left(\mathbf{s}\rightarrow\hat{\mathbf{s}}\right)}
\le P\left(\mathbf{s}\rightarrow\hat{\mathbf{s}}\right)\nonumber\\
    &&\le\underbrace{ C_{U}\left(\ln\rho\right)^{-MN}e^{-\sum_{i=1}^{M}\sum_{j=1}^{N}\frac{\left(\ln \frac{\rho}{\ln^2 \rho} +\ln \left(P_{op}^2\Omega\right)-\ln e_j^2\right)^2}{8\sigma_{ij}^2}}}_{P_{U}\left(\mathbf{s}\rightarrow\hat{\mathbf{s}}\right)}+\mathcal{O}
\left(e^{-\sum_{i=1}^{M}\sum_{j=1}^{N}\frac{\ln^2 \rho}{8\sigma_{ij}^2  }}\right)
\end{eqnarray}
where $\Omega=\sum_{i=1}^{M}\sum_{j=1}^{N}\sigma_{ij}^{-2}$,
 \begin{eqnarray*}
  C_{L}=\frac{\prod_{i=1}^M\prod_{j=1}^N\sigma_{ij}}{\left(4\pi\right)^{MN}\exp\left(-\frac{MN}{2}\right)}Q\left(\frac{1}{2}\left(\sum_{k=1}^Ne_k^2\right)^{-\frac{1}{2}}\right)&& \end{eqnarray*}
  and
  \begin {eqnarray*}
C_{U}=\frac{\left(NP_{op}^2\right)^{MN}}{2\prod_{i=1}^M\prod_{j=1}^N\sqrt{\sigma_{ij}^2}}\exp\left(-\frac{\Omega}{8}\ln^2\left(\frac{NP_{op}^2\Omega}{M}\right)\right)
\end{eqnarray*}~\hfill\QED
\end{theorem}

The detailed proof of Theorem~\ref{theorem:mimo_pep} is provided in Appendix~\ref{app:mimo_pep}.

With all the aforementioned preparations, we are able to give the general design criterion for FDSC of MIMO-OWC in the following subsection.
 \subsection{General Design Criterion for FDSC}

  The discussions in Subsection~\ref{subsec:pep_mimo_ovlc} tells us that $P_{U}\left(\mathbf{s}\rightarrow\hat{\mathbf{s}}\right)$  is the dominant term of the upper-bound of $P\left(\mathbf{s}\rightarrow\hat{\mathbf{s}}\right)$ in \eqref{eqn:pep_asymp}. With this, we will provide a guideline on the space code design in this subsection.  To define the performance parameters to be optimized, we rewrite $P_{U}\left(\mathbf{s}\rightarrow\hat{\mathbf{s}}\right)$ as follows.
\begin{eqnarray}\label{eqn:dominant_term}
P_{U}\left(\mathbf{s}\rightarrow\hat{\mathbf{s}}\right)=
C_{U}\mathcal{G}_{c}\left(\mathbf{e}\right)
\left(\frac{\rho}{\ln^2 \rho}\right)^{\frac{\Omega}{4}\ln\left(
\frac{NP_{op}^2\Omega}{M}\right)-\frac{3}{4}\ln \mathcal{D}_{s}\left(\mathbf{e}\right)}&&\nonumber\\
\times
\left(\ln \rho\right)^{-MN}\exp\left(-\frac{\Omega}{8} \ln^2 \frac{\rho}{\ln^2 \rho}\right)&&
\end{eqnarray}
where $\mathcal{D}_{s}\left(\mathbf{e}\right)=\prod_{j=1}^{N}|e_j|^{\sum_{i=1}^M\sigma_{ij}^{-2}}$ and $\mathcal{G}_{c}\left(\mathbf{e}\right)=\exp\left( \frac{1}{2}\sum_{i=1}^{M}\sum_{j=1}^{N}
\left(\ln |e_j|^{\sigma_{ij}}\right)^2\right)\left(\frac{NP_{op}^2\Omega}{M}\right)^{
\frac{1}{2}\ln\ln \mathcal{D}_{s}\left(\mathbf{e}\right)}$.

Here, the following three factors dictate the minimization of $P_{U}\left(\mathbf{s}\rightarrow\hat{\mathbf{s}}\right)$:
\begin{enumerate}
  \item \textit{Large-scale diversity gain}. The exponent $\Omega$ with respect to $\ln \frac{\rho}{\ln^2 \rho}$ governs the behavior of $P_{U}\left(\mathbf{s}\rightarrow\hat{\mathbf{s}}\right)$. For this reason,  $\Omega$ is named as the \textit{large-scale diversity gain}. The full large-scale diversity achievement is equivalent to the event that all the $MN$ terms in $\Omega=\sum_{i=1}^{M}\sum_{j=1}^{N}\sigma_{ij}^{-2}$ offered by the $N\times M$ MIMO-OWC are fully utilized. On the other hand, $\Omega$ can be considered to be the reciprocal of the channel equivalent variance. In fact, if the channel is independently and identically distributed with variance being $\sigma_{H}^2$,  then $\frac{1}{\Omega}=\frac{\sigma_{H}^2}{MN}$. In this case, the full large-scale diversity gain can be considered to be the reduction of the channel variance by $MN$, which is the maximum reduction amount. In the sense of this point, our definition of $\Omega$ being the large-scale diversity gain is parallel to the diversity order for STBC of MIMO-RF. Thus, when we design space code, full large-scale diversity must be assured \textit{in the first place} to maximize the exponential decaying speed of the error curve.
 \item \textit{Small-scale diversity gain}. $\mathcal{D}_{s}\left(\mathbf{e}\right)=\prod_{j=1}^{N}|e_j|^{\sum_{i=1}^M\sigma_{ij}^{-2}}$ is called \textit{small-scale diversity gain}, which affects the power-law decaying in terms of $\frac{\rho}{\ln^2 \rho}$. $\min_{\mathbf{e}}\mathcal{D}_{s}\left(\mathbf{e}\right)$ should be maximized to optimize the error performance of the worst error event. Since the small-scale diversity gain will affect the average PEP via the power-law decaying speed of the error curve, the small-scale diversity gain of the space code  is what to be optimized \textit{in the second place} to maximize the power-law decaying speed of the error curve.
 \item \textit{Coding gain.} $\mathcal{G}_{c}\left(\mathbf{e}\right)$ is defined as\textit{ coding gain}. On condition that both diversity gains are maximized, if there still exists freedom for further optimization of the coding gain, $\max_{\mathbf{e}\in\mathcal{E}}\mathcal{G}_{c}\left(\mathbf{e}\right)$ should be minimized as the \textit{last step} for the systematical design of space code.
\end{enumerate}

In what follows, we will give a sufficient and necessary condition on a full large-scale diversity achievement. From \eqref{eqn:dominant_term}, we know that  Assumption \ref{assumpt:existence_of_rectangular} is sufficient for FDSC.  Here, from the error detection  perspective, we prove that Assumption \ref{assumpt:existence_of_rectangular} is also necessary. Let us consider the following two possibilities.
\begin{enumerate}
  \item \textit{Bipolar error vector}.  Without loss of generality, assume  there exists an error vector $\mathbf{e}$ containing  bipolar entries, i.e.,  $e_i<0$ and $e_j>0, i\neq j, i,j\in\{1,~\cdots,~N\}$. Then, there  exists a positive vector $\mathbf{h}_0=\left[e_1,~\cdots,~\hat{e}_i,~\cdots,~e_N\right]^T$ such that $\left(\mathbf{e}^T\mathbf{h}_0\right)^2=0$, where $\hat{e}_i=-{\sum_{j=1,j\neq i}^N e_j^2}/{e_i}$.  In addition, for any $\xi>0$, $\mathbf{h}_{\xi}=\xi\mathbf{h}_0$ is positive and $\left(\mathbf{e}^T\mathbf{h}_{\xi}\right)^2=\left(\mathbf{e}^T\mathbf{h}_{0}\xi\right)^2=\left(\xi^2\right)0=0$.
      In other words, there exist two distinct signal vectors $\mathbf{x}$ and $\mathbf{\hat{x}}$ satisfying $\left(\mathbf{x}-\hat{\mathbf{x}}\right)^T\mathbf{h}_0=0$.  That is to say, this signal design
is not able to provide the unique identification of the transmitted signals in the noise-free case,  and, then the reliable detection of the signal will not be guaranteed, even in a sufficiently high SNR.  Therefore, the maximal decaying speed of the corresponding error curve can not be achieved. As a matter of fact, in this case, full large-scale diversity gain can not be attained.
  \item \textit{Error vector with zero entries}.  If $n$ entries of $\mathbf{e}$ are zero, with index being $N-n+1,~\cdots,~N$ without loss of generality, then,  $\sum_{i=1}^M\left(\mathbf{e}^T\mathbf{h}_i\right)^2=\sum_{i=1}^M\left(\sum_{j=1}^{N-n}e_j h_{ij}\right)^2$. Therefore, there must be $Mn$ entries of the channel matrix $\mathbf{H}$ that makes no contribution to PEP. In other words, all the degree of freedoms offered by the $N\times M$ MIMO-OWC are not utilized and, as a consequence,  full large-scale diversity can not be achieved.
\end{enumerate}

To sum up, if either $\mathbf{e}$ is bipolar or has zero-valued entries, the corresponding large-scale diversity gain will be less than $MN$.
 Hence, Assumption \ref{assumpt:existence_of_rectangular} is \textit{sufficient and necessary} for FDSC, which is summarized as the following theorem:

 \begin{theorem} \label{theorem:space_code_full_diversity}
A space code enables full large-scale diversity if and only  if $\forall \mathbf{e}\in \mathcal{E}$, $\mathbf{e}$ is unipolar without zero-valued entries or equivalently, $\forall \mathbf{e}\in \mathcal{E}$, $\mathbf{e}\mathbf{e}^T$ is positive.~\hfill\QED
 \end{theorem}

It is known that for MIMO-RF, the maximal decaying is achieved if and only if the difference matrix of any two distinct codeword matrix is full-rank~\cite{tarokh98}. However, from Theorem~\ref{theorem:space_code_full_diversity}, we can see that a space-only transmission can assure the maximal decaying speed of the corresponding PEP, say, full large-scale diversity gain. The main reason for this significant difference is that the MIMO-OWC channels are nonnegative and the coding matrix is not necessary to be full-rank, which is verified by Theorem.

 Now, the questions raised at the end of Section \ref{sec:model} is answered.
With these results, we can proceed to design FDSC systematically in the following section.

\section{Optimal Design of Specific Linear FDSC }\label{sec:design_example}
In this section, we will exemplify our established criterion in~\eqref{eqn:dominant_term} by designing  a specific \textit{linear} FDSC for $2\times 2$ MIMO-OWC with  unipolar pulse amplitude modulation (PAM). For this particular design,  a closed-form solution to the design problem of space codes optimizing both diversity gains will be obtained by taking advantage of  the available properties on the successive terms of Farey sequences in number theory as well as by developing new properties on the disjoint intervals formed by Farey sequence terms.

\subsection{Design Problem Formulation}
 Consider a $2\times2$ MIMO-OWC system with $\mathbf{F}\left(\mathbf{s}\right)=\mathbf{F}\mathbf{s}$, where
$  \mathbf{F} =
  \left({
  \begin{array}{cc}
  f_{11}&  f_{12}\\
  f_{21}&  f_{22}\\
  \end{array}
  }\right)$ and $\mathbf{e}\mathbf{e}^T=\left(
{\begin{array}{cc}
e_{1}^2&e_1e_2\\
e_1e_2&e_2^2\\
\end{array}
}\right)$.
By Theorem~\ref{theorem:space_code_full_diversity}, $\mathbf{e}\mathbf{e}^T$ should be positive to maximize the large-scale diversity gain.
On the other hand, from  the structure of $\mathbf{e}\mathbf{e}^T$ and~\eqref{eqn:dominant_term},  the small-scale diversity gain is $\mathcal{D}_{s}\left(\mathbf{e}\right)=|e_1e_2|$ under the assumption that CSIT is unknown. Therefore, to optimize the worst case over $\mathcal{E}$, FDSC design is  formulated as follows:
    \begin{eqnarray}\label{eqn:modulator_design}
&&\max_{f_{11},f_{12},f_{21},f_{22}} \min_{\mathbf{e}} e_1e_2\nonumber \\
&& s.t.
\left\{
  \begin{array}{ll}
\left[e_1,e_2\right]^T\in \mathcal{E},f_{ij}>0,i,j\in\{1,2\},\\
e_1e_2>0,f_{11}+f_{12}+f_{21}+f_{22}=1.
  \end{array}
\right.
\end{eqnarray}

Our task is to  analytically  solve \eqref{eqn:modulator_design}.
To do that, we first simplify \eqref{eqn:modulator_design}  by finding all the possible minimum terms.

\subsection{Equivalent  Simplification of Design Problem}\label{subsec:simplification}
For $2^{p}$-PAM, all the possible non-zero values of $e_1e_2$ are
\begin{eqnarray}\label{eqn:objective_function}
e_1e_2=\left(mf_{11}\pm nf_{12}\right)\left(mf_{21}\pm nf_{22}\right)\neq0,m,n\in\mathcal{B}_{2^p}.
\end{eqnarray}
\subsubsection{Preliminary simplification}
After observations over  \eqref{eqn:objective_function},  we have the following  facts.
 \begin{enumerate}
   \item  $\forall m\neq0,m,n\in\mathcal{B}_{2^p}$, it holds that
   \begin{eqnarray*}
\left(mf_{11}+ nf_{12}\right)\left(mf_{21}+nf_{22}\right)
  \ge f_{11}f_{21}.
   \end{eqnarray*}
   \item $\forall n\neq0$, $m,n\in\mathcal{B}_{2^p}$, it is true that
   \begin{eqnarray*}
\left(mf_{11}+ nf_{12}\right)\left(mf_{21}+nf_{22}\right)\ge f_{12}f_{22}.
   \end{eqnarray*}
   \item $\forall k\neq0,m^2+n^2\neq0,k,m,n\in\mathcal{B}_{2^p}$, we have
   \begin{eqnarray*}
\frac{k\left(mf_{11}-  nf_{12}\right)\left( mf_{21}- nf_{22}\right)}{\left(mf_{11}-nf_{12}\right)\left(mf_{21}-nf_{22}\right)}
   \ge 1.
   \end{eqnarray*}
 \end{enumerate}
So,  all the possible minimum of $e_1e_2$ in  \eqref{eqn:modulator_design}  are
$f_{11}f_{21}$, $f_{12}f_{22}$ and $\left(mf_{11}-nf_{12}\right)\left(mf_{21}-nf_{22}\right)$,
where $\frac{n}{m}$ is irreducible, i.e., $m\perp n$. These terms are denoted by
\begin{eqnarray}
&&F_{10}=f_{12}f_{22}\left(\frac{f_{11}}{f_{12}}\times\frac{f_{21}}{f_{22}}\right),F_{01}=f_{12}f_{22},
\nonumber\\
&&F_{mn}=f_{12}f_{22}\left(m\frac{f_{11}}{f_{12}}-n\right)\left(m\frac{f_{21}}{f_{22}}-n\right).\nonumber
\end{eqnarray}
After putting aside the common term, $f_{12}f_{22}$, we can see that $F_{mn}$ is the piecewise linear function of $\frac {f_{11}}{f_{12}}$ and $ \frac{f_{21}}{f_{22}}$, respectively. So,  \eqref{eqn:modulator_design} can be solved by fragmenting interval $\left[0,\infty\right)$ into disjoint subintervals. This fragmentation can be done by  the breakpoints where $F_{mn}=0$.
To  characterize this sequence, there exists an elegant mathematical tool in number theory presented below.
   \subsubsection{Farey sequences}
First, we observe some specific examples of the breakpoint sequences.
For OOK, the breakpoints $\frac{0}{1}, \frac{1}{1},\infty$.
For 4-PAM, they are $\frac{0}{1},\frac{1}{3},\frac{1}{2},\frac{2}{3},\frac{1}{1},\frac{3}{2},\frac{2}{1},\frac{3}{1},\infty$.
 For  8-PAM, we have the breakpoint sequence with the former part being
   \begin{subequations}
   \begin{eqnarray}\label{eqn:before_1}
\frac{0}{1},\frac{1}{7},\frac{1}{6},\frac{1}{5},\frac{1}{4},
\frac{2}{7},\frac{1}{3},\frac{2}{5},\frac{3}{7},\frac{1}{2},
\frac{4}{7},\frac{3}{5},\frac{2}{3},\frac{5}{7},
  \frac{3}{4},\frac{4}{5},\frac{5}{6},\frac{6}{7},\frac{1}{1}&&
  \end{eqnarray}
and the remaining being
  \begin{eqnarray}\label{eqn:after_1}
 \frac{7}{6},\frac{6}{5},\frac{5}{4},\frac{4}{3}, \frac{7}{5},\frac{3}{2},\frac{5}{3},\frac{7}{4},\frac{2}{1},
  \frac{7}{3},\frac{5}{2},\frac{3}{1},\frac{7}{2},\frac{4}{1},\frac{5}{1},\frac{6}{1}
              ,\frac{7}{1},\infty&&
   \end{eqnarray}
   \end{subequations}
  Through  these special examples, we find that the series of breakpoints before $1/ 1$ (such as the sequence in \eqref{eqn:before_1}) is
the one which is called the Farey sequence~\cite{hardy1979introduction}.
The Farey sequence $\mathfrak{F}_k$ for any positive integer $k$ is the set
 of irreducible rational numbers $\frac{a}{b}$ with $0\leq a\leq b\leq k$ arranged in an increasing order.
 The series of breakpoints after $\frac{1}{1}$ (such as the sequence in \eqref{eqn:after_1}) is
the reciprocal version of the Farey sequence. Thus, our  focus is on the sequence before $\frac{1}{1}$.

The Farey sequence has many interesting properties~\cite{hardy1979introduction}, some of which closely relevant to our problem are given as follows.
\begin{proposition}\label{lemma:farey_sequence}
If $\frac{n_1}{ m_1}$, $\frac{n_2}{ m_2}$ and $\frac{n_3}{ m_3}$ are three successive terms of $\mathfrak{F}_k,k>3$ and $\frac{n_1}{ m_1}<\frac{ n_2}{ m_2}<\frac{ n_3}{ m_3}$,
then,
\begin{enumerate}
  \item $ m_1n_2-m_2n_1=1$ and $m_1+m_2\ge k+1$.
  \item $\frac{n_1+n_2}{m_1+m_2}\in\left(\frac{n_1}{m_1},\frac{n_3}{m_3}\right)$ and $\frac{n_2}{m_2}=\frac{n_1+n_3}{m_1+m_3}$.~\hfill\QED
\end{enumerate}

\end{proposition}

However, having only Proposition~\ref{lemma:farey_sequence} is not enough to solve our  design problem in \eqref{eqn:modulator_design}. We need to develop the other important new properties of Farey sequences, which will be
utilized in the FDSC design in the ensuing subsections.

\begin{property} \label{property:local_two_worst_cases} Given $k>3$, assume $\frac{n_0}{ m_0},\frac{n_1}{m_1},\frac{n_2}{ m_2},\frac{n_3}{ m_3}\in\mathfrak{F}_{k}$ and $\frac{n_0}{m_0}<\frac{n_1}{m_1}<\frac{n_2}{m_2}<\frac{n_3}{m_3}$.
If $\frac{n_1}{m_1}$ and $\frac{n_2}{ m_2}$  are successive,
then, $\frac{n_1+n_3}{m_1+m_3}\ge\frac{n_2}{m_2}$ and $\frac{n_0+n_2}{m_0+m_2}\le\frac{n_1}{m_1}$.
~\hfill\QED
\end{property}

The proof of Property~\ref{property:local_two_worst_cases} is postponed into Appendix~\ref{app:local_last_two_worst_cases}.

\begin{property} \label{property:the_local_solution}
Assume $\frac{n_1}{m_1},\frac{n_2}{m_2}\in \mathfrak{F}_{k}, k>3$ and $\frac{n_1}{m_1}<\frac{n_2}{m_2}$.   Then,
  \begin{enumerate}
     \item $\frac{n_1}{m_1}<\frac{n_1+n_2}{m_1+m_2}<\frac{n_2}{m_2}$ holds.
      \item If $\frac{f_{11}}{f_{12}},\frac{f_{21}}{f_{22}}\in\left(\frac{n_1}{m_1},\frac{n_1+n_2}{m_1+m_2} \right)$, then,  $F_{m_1n_1}<F_{m_2n_2}$.
      \item If $\frac{f_{11}}{f_{12}},\frac{f_{21}}{f_{22}}\in\left(\frac{n_1+n_2}{m_1+m_2},\frac{n_2}{m_2} \right)$, then,  $F_{m_1n_1}>F_{m_2n_2}$.
      \item If $\frac{f_{11}}{f_{12}}=\frac{f_{21}}{f_{22}}=\frac{n_1+n_2}{m_1+m_2}$, then, $F_{m_1n_1}=F_{m_2n_2}$.~\hfill\QED
    \end{enumerate}
\end{property}

The proof of Property~\ref{property:the_local_solution} is given in Appendix~\ref{app:the_local_solution}.

Using Properties~\ref{property:local_two_worst_cases} and~\ref{property:the_local_solution}, we attain the following property.
\begin{property}\label{th:min}
If $\frac{n_1}{m_1}$ and $\frac{n_2}{m_2}$ are successive in $\mathfrak{F}_{k}$ and $\frac{f_{11}}{f_{12}},\frac{f_{21}}{f_{22}}\in\left(\frac{n_1}{m_1},\frac{n_2}{m_2} \right)$,
then, $  F_{m_1n_1}$ and $F_{m_2n_2}$ are the two worst cases.~\hfill\QED
\end{property}

The proof of Property~\ref{th:min} is provided in Appendix~\ref{app:th:min}.

\subsection{Techniques to Solve The Max-min Problem}\label{subsec:max_min}
  Thanks to Farey sequences, \eqref{eqn:modulator_design} is transformed into a piecewise max-min problem  with two objective functions. By solving this kind of problem, another main result of this paper can be formally presented as the following theorem.
\begin{theorem}\label{theorem:golbal_solution}
The solution  to~\eqref{eqn:modulator_design} is determined by
 \begin{subequations}\label{eqn:global_optimal_modulator}
\begin{eqnarray}
  \mathbf{F}_{FDSC} =\frac{1}{2+2^{p+1}}\left(
  {\begin{array}{ccc}
  1&2^p\\
  1&2^p\\
 \end{array}}
  \right),
  \end{eqnarray}
 or
\begin{eqnarray}
  \mathbf{F}_{FDSC}=\frac{1}{2+2^{p+1}}\left(
  {\begin{array}{ccc}
  2^p&1\\
  2^p&1\\
 \end{array}}
  \right).
\end{eqnarray}
\end{subequations}
~\hfill \QED
\end{theorem}

The proof of Theorem \ref{theorem:golbal_solution} is postponed into Appendix \ref{app:global_solution}.

Theorem~\ref{theorem:golbal_solution} uncovers the fact that the optimal linear space coded symbols are actually unipolar $2^{2p}$-ary PAM symbols,
since $\mathcal{B}_{2^{2p}}=\{s_1+2^p s_2:s_1,s_2\in\mathcal{B}_{2^p}\}$. Therefore, in fact, we have proved that RC~\cite{navidpour2007itwc} is optimal in the sense of the criterion established in this paper.

\section{Computer Simulations }\label{sec:numerical_results}
\begin{figure}[!htp]
    \centering
    \resizebox{7cm}{!}{\includegraphics{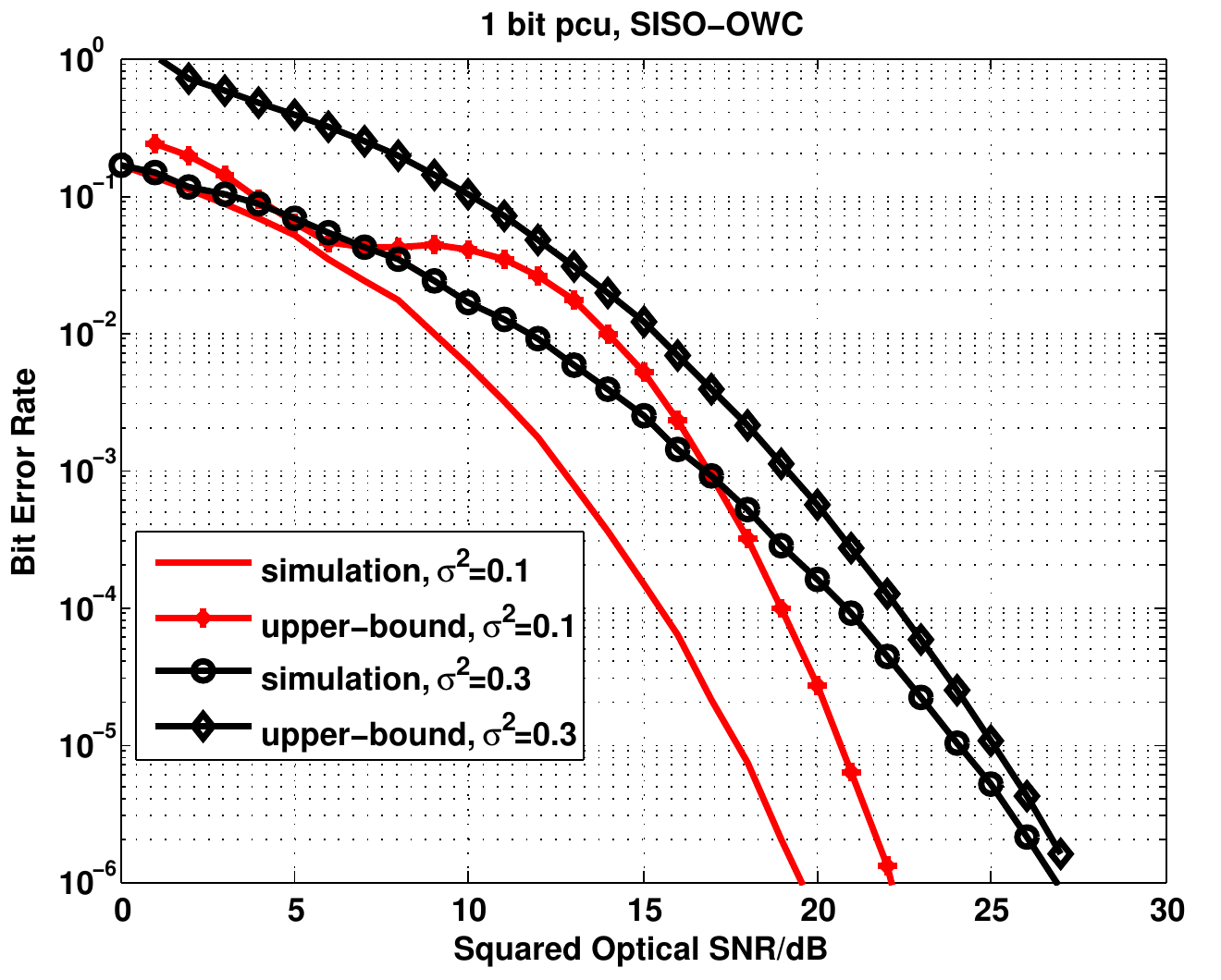}}
    \centering \caption{Error performance of SISO-OWC }
    \label{fig:siso}
\end{figure}

\begin{figure}[!htp]
    \centering
    \resizebox{7cm}{!}{\includegraphics{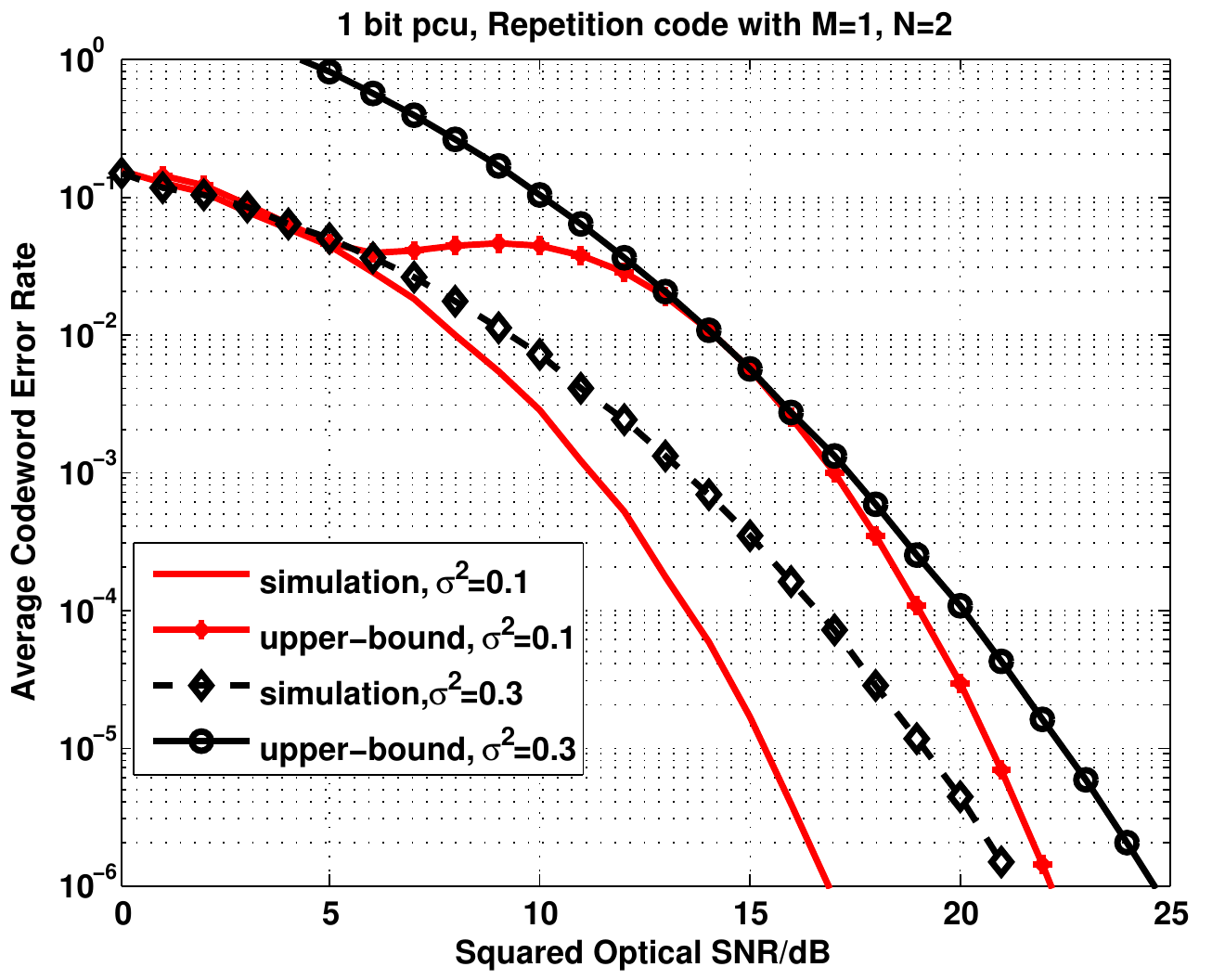}}
    \centering \caption{Error performance of MIMO-OWC }
    \label{fig:mimo}
\end{figure}

In this section, we carry out computer simulations to verify our theoretical results. We first examine the performance bounds given in Lemma 1 and Theorem 1.  As shown by Figs.~\ref{fig:siso} and ~\ref{fig:mimo}, it can be seen that in the high SNR regimes, the proposed upper-bounds have almost the same negative slope as those of simulated results. In other words, the proposed upper-bounds have captured the behavior of the error curve with respect to the decaying speeds. As shown by Figs.~\ref{fig:siso} and~\ref{fig:mimo}, when SNR is sufficiently high, our proposed bounds have a horizontal shift to the right compared with the simulated results. Therefore, the tightness of the proposed asymptotical bounds is mainly dependent on the precise estimate of  some constant independent of SNR.


In the following, we simulate to verify our newly developed criterion in \eqref{eqn:dominant_term}.
 In light of our work being initiative, the only space-only transmission scheme available in the literature is spatial multiplexing. Accordingly,  we compare the performance of spatial multiplexing and FDSC specifically designed for $2\times2$ MIMO-OWC in Section \ref{sec:design_example}.
In addition, we suppose that $h_{ij},i,j=1,2$ are independently and identically distributed and let $\sigma_{11}=\sigma_{12}=\sigma_{21}=\sigma_{22}=\sigma$.
 These schemes are  as follows:
 \begin{enumerate}
  \item \textit{FDSC}.  The optical power is  normalized in such a way that $\sum_{i,j=1}^2f_{ij}=2$ yields $E\left[\sum_{i,j=1}^2 f_{ij}s_j\right]=1$. From  \eqref{eqn:global_optimal_modulator}, the coding matrix is
$  \mathbf{F}_{FDSC} =
  \frac{1}{3}\left({
  \begin{array}{cc}
  2&  1\\
  2&  1\\
  \end{array}
  }\right)$.
  \item \textit{Spatial Multiplexing}.  We fix the modulation formats to be OOK and vary $\sigma^2$. So the rate is 2 bits per channel use (pcu). The  transmitted symbols $s_1,s_2$ are chosen from $\{0,1\}$ equally likely. The average optical power is $E\left[s_1+s_2\right]=1$.
  \end{enumerate}
\begin{figure}[!htp]
    \centering
    \resizebox{7cm}{!}{\includegraphics{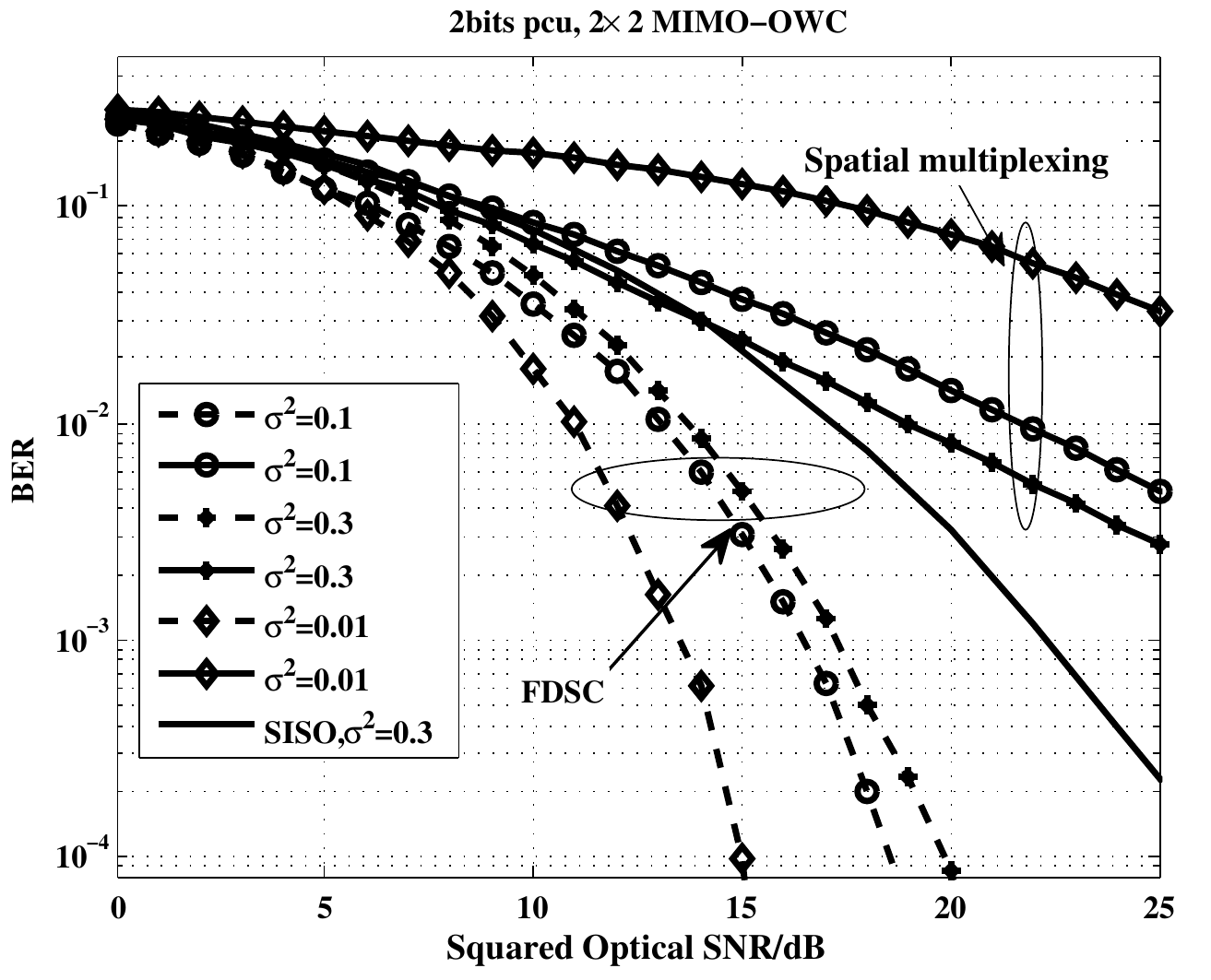}}
    \centering \caption{BER  comparisons of  FDSC and spatial multiplexing.}
    \label{fig:modulated_unmodulated}
\end{figure}
\begin{figure}[!htp]
    \centering
    \resizebox{7cm}{!}{\includegraphics{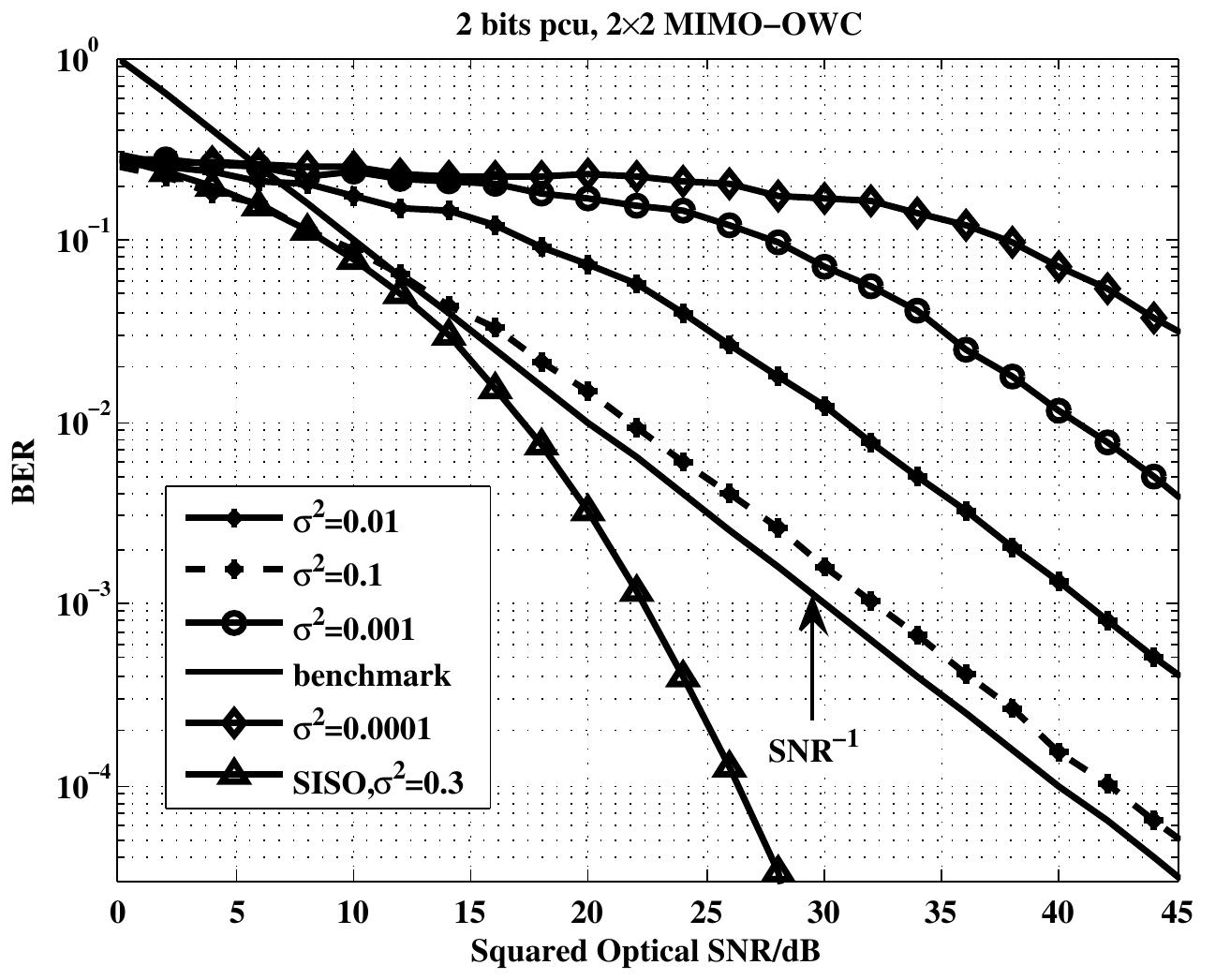}}
    \centering \caption{BER comparisons of spatial multiplexing.}
    \label{fig:multiplexing_MIMO}
\end{figure}
 We can see that both schemes  have the same spectrum efficiency, i.e., 2 bits pcu and the same optical power. Through numerical results, we have following observations.
\begin{enumerate}
  \item Substantial enhancement from FDSC is achieved, as shown in Fig.~\ref{fig:modulated_unmodulated}. For $\sigma^2=0.01$, the improvement is almost 16 dB at the target bit error rate (BER) of $10^{-2}$. For $\sigma^2=0.5$, the improvement is almost 6 dB at the target BER of $10^{-3}$. Note that the small-scale gain also governs the negative slope of error curve. The decaying speed of the error curve of FDSC is exponential in terms of $\ln\frac{\rho}{\ln^2\rho}$ and that of spatial multiplexing is power-law with respect to $\rho$. The reason for this difference is that FDSCs are full-diversity guaranteed by the positive constraints in \eqref{eqn:modulator_design}, whereas spatial multiplexing does not satisfy the positive requirement in Theorem \ref{theorem:space_code_full_diversity}.
\item Spatial multiplexing presents only small-scale diversity gain illustrated in Fig. \ref{fig:multiplexing_MIMO}.  By varying the variance of $\mathbf{H}$, we find that in the high SNR regimes, the error curve decays as $\rho^{-1}$ as long as the SNR is high enough. From $\sigma^2=0.001$ to $\sigma^2=0.1$, the error curve has a horizonal shift, which is the typical style of MIMO RF~\cite{tarokh98}. The reason is given as follows. The equivalent space coding matrix is
      $\mathbf{e}^T\mathbf{e}=\left(
{\begin{array}{cc}
e_{1}^2&e_1e_2\\
e_1e_2&e_2^2\\
\end{array}
}\right),e_1,e_2\in\{0,\pm1\}$ with $e_1^2+e_2^2\neq0$. It should be noted that there exists two typical error events: $e_1e_2=-1$ and $e_1e_2=0$. From the necessity proof of Theorem \ref{theorem:space_code_full_diversity}, for $e_1e_2=-1$, the attained large-scale diversity gain is zero, and at the same time, if $e_1e_2=0$ with $e_1^2+e_2^2\neq0$, then, the attained large-scale diversity gain is only two for $2\times2$  MIMO-OWC. Therefore, the overall large-scale diversity gain of spatial multiplexing is zero with small-scale diversity gain being attained.
\item  From Fig.~\ref{fig:modulated_unmodulated} and~\ref{fig:multiplexing_MIMO}, we notice that when $\sigma$ increases, the error performances of repetition codes will worsen. For this phenomenon, the reason is that for repetition codes, the large-scale diversity is given by $\Omega=\sum_{i,j=1}^2\sigma_{ij}^{-2}$, which dominates the exponential decaying speed. However, for spatial multiplexing, increasing $\sigma^2$ will improves the error performance by providing a horizontal shift to the left. It is known that since an exact error probability formula for OWC over log-normal fading channels is indeed hard to be obtained particularly for MIMO-OWC, it is very challenging to theoretically prove that the relationship between the error performance of spatial multiplexing and $\sigma^2$. Intuitively speaking, the relationship between the error performance of spatial multiplexing and $\sigma^2$ follows the radio frequency MIMO. That is to say, $\sigma^2$ is the average electrical power of $\ln h$. That's why increasing $\sigma^2$ will improve the error performance of spatial multiplexing, as shown by Figs.~\ref{fig:modulated_unmodulated} and~\ref{fig:multiplexing_MIMO}.
\end{enumerate}

We can see that the performance gain of MIMO-OWC if any  will become larger against increasing SNR in a high enough regime. This implies that a slight improvement in the coding structure will result in a significant enhancement of error performance in the high SNR regimes instead of only an horizonal shift to the left. Unique characteristics of  MIMO-OWC are experimentally uncovered and our established criterion are verified.

\section{Conclusion and Discussions}
In this paper, we have established a general criterion on the full-diversity  space coded transmission of MIMO-OWC for the ML receiver, which is, to our best knowledge, the first design criterion for the full-diversity transmission of optical wireless communications with IM/DD over log-normal fading channels. Particularly for a $2\times 2$ case, we have attained a closed-form solution to the optimal linear FDSC design problem, proving that RC is the optimal among all the linear space codes. Our results indicate that the transmission design is indeed necessary and essential for significantly improving overall error performance for MIMO-OWC. However, the design criterion and the specific code constructions for MIMO-OWC presented in this paper are just initiative. Some significant issues are under consideration:
 \begin{enumerate}
   \item \textit{Our proposed criterion can be applicable to any non-linear designs. It remains open whether  there exists any better non-linear space code than  RC}.
   \item \textit{It has been shown in this paper that the space dimension  alone is sufficient for full large-scale diversity. Here, a natural question is: what kind of benefit can be obtained if space-time block code designs are used for MIMO-OWC}?
   \item \textit{Like MIMO techniques for RF communications, what is the diversity-multiplexing tradeoff for MIMO-OWC?}
 \end{enumerate}

\appendix
\subsection{ Proof of Lemma~\ref{lemma:siso_sep}}\label{app:siso_sep}
 Our main idea here is to split the whole integral in~\eqref{eqn:ser-siso} into two parts by properly choosing $\tau>0$ such that its dominant term can be extracted. In other words, the average SEP~\cite{Proakis00} can be rewritten by
 \begin{eqnarray} \label{eqn:ber_siso_segment}
  P_e\left(\rho\right)&=&\int_{0}^{\tau}Q\left(\frac{\sqrt{\rho }h}{P_{op}}\right)f_{H}\left(h\right) dh \nonumber\\
   &+&\int_{\tau}^{\infty}Q\left(\frac{\sqrt{\rho }h}{P_{op}}\right)f_{H}\left(h\right) dh.
\end{eqnarray}
 Now, we select $\tau$ to satisfy that when $\rho\rightarrow\infty$, $\tau\rightarrow0$, to fragment $\left(0,\infty\right)$ into $\left(0,\tau\right)$ and $\left(\tau,\infty\right)$, adaptively with SNR. This fragmentation is to find the dominant term of $ P_e\left(\rho\right)$ in the high SNR regimes by giving the upper-bound and the lower bound of $\int_{0}^{\tau}Q\left(\frac{\sqrt{\rho}h}{P_{op}}\right)f_{H}\left(h\right) dh $, and the upper-bounds of $\int_{\tau}^{\infty}Q\left(\frac{\sqrt{\rho}h}{P_{op}}\right)f_{H}\left(h\right) dh $ and then, examining their asymptotical behaviors related with $\tau$. Temporarily, we take this fragmentation for granted and then, will explain the essential reason later on.
\subsubsection{Upper-bound of SEP over $\left(0,\tau\right)$}
We integrate the first part of SEP in~\eqref{eqn:ber_siso_segment}, which is denoted by ${P}_{\tau}\left(\rho\right)$. Notice that when $\rho\rightarrow\infty$, $\tau\rightarrow0$, and in this case,  $f_{H}\left(h\right)$ is  monotonically increasing over $\left(0,\tau\right)$.
Hence, 
${P}_{\tau}\left(\rho\right)$ is upper-bounded by
 \begin{eqnarray} \label{eqn:ber_siso_upper_bound1}
{P}_{\tau}\left(\rho\right)&\le &f_{H}\left(\tau\right) \int_{0}^{\tau}Q\left(\frac{\sqrt{\rho }h}{P_{op}}\right)dh \nonumber \\
&\le&  \frac{f_{H}\left(\tau\right)}{2}\int_{0}^{\tau}\exp\left(-\frac{\rho h^2}{2P_{op}^2}\right)dh
 \end{eqnarray}
where the last inequality is obtained by using the Chernoff bound on the Gaussian tail integral.
Furthermore, we can upper-bound the last integral of \eqref{eqn:ber_siso_upper_bound1} by
\begin{eqnarray}
\int_{0}^{\tau}\exp\left(-\frac{\rho h^2}{2P_{op}^2}\right)dh&=&P_{op}\sqrt{2\pi}\rho^{-\frac{1}{2}} \left(\frac{1}{2}-Q\left(\frac{\sqrt{{\rho\tau^2 }}}{P_{op}}\right)\right)\nonumber\\
&\le& \frac{P_{op}\sqrt{2\pi}}{2}\rho^{-\frac{1}{2}}
\end{eqnarray}
which produces
  \begin{eqnarray} \label{eqn:ber_siso_upper_bound}
{P}_{\tau}\left(\rho\right)&\le&
\frac{P_{op}\sqrt{2\pi}f_{H}\left(\tau\right)}{4}\rho^{-\frac{1}{2}}\nonumber\\
&=&\frac{P_{op}\exp\left(\frac{\sigma^2}{2}\right)}{4\sqrt{\sigma^2}}
\rho^{-\frac{1}{2}}
\exp\left(-\frac{\left(\ln \tau+\sigma^2\right)^2}{2\sigma ^{2}}\right)
\end{eqnarray}

 \subsubsection{Lower-bound of SEP}
At the same time, when $\rho\rightarrow\infty$, we have $\frac{\tau}{\ln \rho}\le \tau$.
This inequality allows us to lower-bound ${P}_{\tau}\left(\rho\right)$ by
  \begin{eqnarray}
{P}_{\tau}\left(\rho\right)&=&\int_{0}^{\tau}Q\left(\frac{\sqrt{\rho }h}{P_{op}}\right)f_{H}\left(h\right) dh \nonumber \\
&\ge&\int_{0}^{\frac{\tau}{\ln \rho}}Q\left(\frac{\sqrt{\rho }h}{P_{op}}\right)f_{H}\left(h\right) dh \nonumber \\
&\ge&\int_{0}^{\frac{\tau}{\ln \rho}}Q\left(\frac{\sqrt{\rho }\tau}{P_{op}\ln \rho}\right)f_{H}\left(h\right) dh
 \end{eqnarray}
 where the last inequality follows from the monotonically decreasing property of $Q$-function.

 Then, by integrating $f_{H}\left(h\right)$ over $\left(0,\frac{\tau}{\ln \rho}\right)$, we have
   \begin{eqnarray}\label{eqn:ber_siso_lower_bound}
&&{P}_{\tau}\left(\rho\right)\ge Q\left(\frac{\sqrt{\rho }\tau}{P_{op}\ln \rho}\right) Q\left(\frac{-\ln \tau+\ln \ln \rho}{\sigma} \right) \nonumber \\
&&\ge \frac{Q\left(\frac{\sqrt{\rho }\tau}{P_{op}\ln \rho}\right)}{2\sqrt{2\pi}}\frac{\sigma}{\ln \frac{ \ln \rho }{\tau}}\exp\left(-\frac{\left(\ln \frac{\tau}{ \ln \rho }\right)^2}{2\sigma^2}\right)
 \end{eqnarray}
 where the last inequality is obtained by using
 \begin{eqnarray}\label{eqn:q_function_lower_bound}
Q\left(x\right)\ge \frac{1}{2\sqrt{2\pi}x}\exp\left(-\frac{x^2}{2}\right),x\ge\frac{\sqrt{2}}{2}.
\end{eqnarray}

\subsubsection{Upper-bound of SEP over $\left(\tau,\infty\right)$}
 We now turn to the second part of $P_e\left(\rho\right)$ in~\eqref{eqn:ber_siso_segment}, which is denoted by $\bar{P}_\tau\left(\rho\right)$.
Letting $f'_{H}\left(h\right)=0$ produces the point $h_0$ satisfying $\forall h\ge0,f_{H}\left(h\right)\le f_{H}\left(h_0\right)=\frac{\exp\left(\frac{\sigma^2}{2}\right)}{\sqrt{2\pi\sigma^2}}$. With this,  $\bar{P}_{\tau}\left(\rho\right)$ can be upper-bounded by
   \begin{eqnarray*}
\bar{P}_\tau\left(\rho\right)
&\le& f_{H}\left(h_0\right)\int_{\tau}^{\infty}Q\left(\frac{\sqrt{\rho }h}{P_{op}}\right) dh
 \end{eqnarray*}
 Using Chernoff bound on the Gaussian tail integral gives us
\begin{eqnarray}\label{eqn:pep-siso-up}
\bar{P}_\tau\left(\rho\right)&\le&
\frac{1}{2}f_{H}\left(h_0\right)\int_{\tau}^{\infty}\exp\left(-\frac{\rho h^2}{2P_{op}^2}\right) dh\nonumber \\
&=&\frac{1}{2}f_{H}\left(h_0\right)\sqrt{2\pi}\rho^{-1/2}Q\left(\frac{\sqrt{{\rho\tau^2 }}}{P_{op}}\right) \nonumber \\
&\le&\frac{\exp\left(\frac{\sigma^2}{2}\right)}{4\sqrt{\sigma^2}}\rho^{-1/2}\exp\left(-\frac{\rho\tau^2}{2P_{op}^2}\right)
 \end{eqnarray}
\subsubsection{Determination of $\tau$}
Thus far, we have attained three bounds, respectively shown in  \eqref{eqn:ber_siso_upper_bound}, \eqref{eqn:ber_siso_lower_bound} and \eqref{eqn:pep-siso-up}. Now, we examine their tightness related to $\tau$.
We can see that ~\eqref{eqn:ber_siso_upper_bound} and \eqref{eqn:ber_siso_lower_bound}  have the same exponential term $\frac{\left(\ln \tau\right)^2}{\sigma^2}$. To make the upper bounds in \eqref{eqn:ber_siso_upper_bound} to approach the upper-bound in \eqref{eqn:pep-siso-up} as tightly as possible, we seek to select such $\tau$ that
\begin{eqnarray}\label{eqn:hyper_equation_siso}
\frac{\rho\tau^2}{2P_{op}^2}=\frac{\left(\ln \tau+\sigma^2\right)^2}{2\sigma ^{2}}
\end{eqnarray}
An explicit solution to~\eqref{eqn:hyper_equation_siso} with respect to $\tau$  is difficult to obtain. Accordingly, we propose to approximate  the solution to \eqref{eqn:hyper_equation_siso} by
 \begin{eqnarray}\label{eqn:tau_selection_siso}
\tau=\frac{P_{op}\ln \rho}{2\sqrt{\rho\sigma^{2}}}
 \end{eqnarray}
 which satisfies the following asymptotical equality.
 \begin{eqnarray*}
\lim_{\rho\rightarrow\infty}\frac{\frac{\rho\tau^2}{2P_{op}^2}}{\frac{\left(\ln \tau+\sigma^2\right)^2}{2\sigma ^{2}}}=\lim_{\rho\rightarrow\infty} \frac{\left(\ln\rho\right)^2}{\left(\ln\frac{\rho}{\ln^2\rho}-\ln\frac{ P^2_{op}}{4\sigma^2}\right)^2}=1
 \end{eqnarray*}

Finally, combining \eqref{eqn:ber_siso_upper_bound}, \eqref{eqn:ber_siso_lower_bound}, \eqref{eqn:pep-siso-up}, and \eqref{eqn:tau_selection_siso}  leads to the fact that there exist  three positive constants $C_{1L}$, $C_{11U}$ and $C_{12U}$ independent of $\rho$ such that
  \begin{eqnarray}\label{eqn:ber_siso}
  &&C_{1L}\frac{1}{\ln\rho-\ln \frac{4\sigma^{2}}{P_{op}^2}}\exp\left(-\frac{\left(\ln\rho -\ln\frac{P_{op}^2}{4\sigma^{2}}\right)^2}{8\sigma^2}\right)
  \le P_e(\rho)\nonumber\\
  &&\le \frac{1}{4}\left(\ln\rho\right)^{-1}  \exp\left(-\frac{\left(\ln\frac{\rho}{\ln^2\rho}-\ln\frac{ P^2_{op}}{4\sigma^2}\right)^2}{8\sigma ^2}\right)
  \nonumber\\
  &&+C_{12U}
  \rho^{-\frac{1}{2}}\exp\left(-\frac{\left(\ln\rho\right)^2}{8\sigma^2}\right)
  \end{eqnarray}
where  $C_{1L}=\frac{Q\left(\frac{1}{2\sigma}\right)}{\sqrt{2\pi}}$, $C_{11U}=\frac{1}{2}$ and $C_{12U}=\frac{\exp\left(\frac{\sigma^2}{2}\right)}{4\sqrt{\sigma^2}}$.
This is the complete proof of Lemma~\ref{lemma:siso_sep}.~\hfill$\Box$

\subsection{Proof of Theorem~\ref{theorem:mimo_pep}}\label{app:mimo_pep}
 The condition that any $\mathbf{e}\in {\mathcal E}$ is unipolar without zero entry implies that  if $\forall \tau>0$ and $\forall \mathbf{e}\in {\mathcal E}$, $\left(\mathbf{e}^T\mathbf{h}_{i}\right)^2\le \tau^2$, then, we can have $h_{ij}\le\frac{\tau}{|e_j|},i=1,\ldots,M,j=1,\ldots,N$.
 After these preparations, we adopt the same techniques as Subsection \ref{subsec:siso_sep}. Temporally, assume that when $\rho\rightarrow \infty$, $\tau\rightarrow 0$.  Similarly, $P\left(\mathbf{s}\rightarrow\hat{\mathbf{s}}\right)$ can be adaptively fragmented with SNR as
\begin{eqnarray} \label{eqn:ml_detection_pep_fragment}
P\left(\mathbf{s}\rightarrow\hat{\mathbf{s}}\right)
=\int_{\left(\mathbf{e}^T\mathbf{h}_j\right)^2\le\tau^2}P\left(\mathbf{s}\rightarrow\hat{\mathbf{s}}|\mathbf{H}\right)f_{\mathbf{H}}\left(\mathbf{H}\right)d\mathbf{H}
+\int_{ \left(\mathbf{e}^T\mathbf{h}_j\right)^2>\tau^2}P\left(\mathbf{s}\rightarrow\hat{\mathbf{s}}|\mathbf{H}\right)f_{\mathbf{H}}\left(\mathbf{H}\right)d\mathbf{H}.
\end{eqnarray}

The target in the ensuing subsections is to give the asymptotical bounds on  $P\left(\mathbf{s}\rightarrow\hat{\mathbf{s}}\right)$ following the similar procedures to the case of SISO-OWC.
\subsubsection{Upper-bound of PEP over $\left(0,\tau\right)$}
To begin with, let us  process the first part of $P\left(\mathbf{s}\rightarrow\hat{\mathbf{s}}\right)$ in \eqref{eqn:ml_detection_pep_fragment} denoted by $P_{\tau}\left(\mathbf{s}\rightarrow\hat{\mathbf{s}}\right)$.
 We know when $\rho\rightarrow\infty$, $\tau\rightarrow0$.  In this instance, $f_{H}\left(h_{ij}\right)$ is  monotonically increasing over $\left(0,\frac{\tau}{|e_i|}\right)$ and then, $f_{H}\left(h_{ij}\right)\le f_{H}\left(\frac{\tau}{|e_j|}\right)$.
Together with the Chernoff bound of $Q$-function,  $P_{\tau}\left(\mathbf{s}\rightarrow\hat{\mathbf{s}}\right)$ can be upper-bounded by
\begin{subequations}
\begin{eqnarray}\label{eqn:pep_upper_tau}
P_{\tau}\left(\mathbf{s}\rightarrow\hat{\mathbf{s}}\right)\le \frac{1}{2}\prod_{j=1}^N\prod_{i=1}^Mf_{H_{ij}}\left(\frac{\tau}{|e_j|}\right)
\int_{\left(\mathbf{e}^T\mathbf{h}_i\right)^2\le\tau^2}
\exp\left(-\frac{\rho\sum_{i=1}^M\left(\mathbf{e}^T\mathbf{h}_i\right)^2}{8NP_{op}^2}\right)d\mathbf{H} \end{eqnarray}
In addition, by Assumption \ref{assumpt:existence_of_rectangular}, $e_1,~\cdots,~e_N$ have the same signs. This result allows us to further upper-bound $P_{\tau}\left(\mathbf{s}\rightarrow\hat{\mathbf{s}}\right)$ by
\begin{eqnarray}\label{eqn:pep_upper_Q_function}
P_{\tau}\left(\mathbf{s}\rightarrow\hat{\mathbf{s}}\right)\le \frac{1}{2}\prod_{j=1}^N\prod_{i=1}^Mf_{H_{ij}}\left(\frac{\tau}{|e_j|}\right)
\times\int_{h_{ij}\le \frac{\tau}{|e_j|}}
\exp\left(-\frac{\rho\sum_{j=1}^N\sum_{i=1}^Me_j^2h_{ij}^2}{8NP_{op}^2}\right)d\mathbf{H}
\end{eqnarray}
Further, integrating the last term in \eqref{eqn:pep_upper_Q_function}  produces
\begin{eqnarray}\label{eqn:pep_upper_bound}
&&P_{\tau}\left(\mathbf{s}\rightarrow\hat{\mathbf{s}}\right)\le \frac{1}{2}\prod_{j=1}^N\prod_{i=1}^Mf_{H_{ij}}\left(\frac{\tau}{|e_j|}\right)
\prod_{j=1}^N\prod_{i=1}^M\sqrt{2\pi}
\left(\frac{\rho|e_j|^2}{4NP_{op}^2}\right)^{-\frac{1}{2}}
\left(\frac{1}{2}-Q\left(\frac{\rho\tau^2}{4NP_{op}^2}\right)\right)\nonumber \\
&&
\le \frac{1}{2}\prod_{j=1}^N\prod_{i=1}^M\frac{\sqrt{2\pi}}{2}f_{H_{ij}}\left(\frac{\tau}{|e_j|}\right)
\left(\frac{\rho|e_j|^2}{4NP_{op}^2}\right)^{-\frac{1}{2}}
\nonumber \\
&&
= \frac{\tau^{-MN}}{2\prod_{j=1}^N\prod_{i=1}^M\sqrt{\sigma_{ij}^2}}
\left(\frac{\rho}{NP_{op}^2}\right)^{-\frac{MN}{2}}
\exp\left(-\sum_{j=1}^N\sum_{i=1}^M\frac{\left(\ln\tau-\ln|e_j|\right)^2}{2\sigma_{ij}^2}\right)
\end{eqnarray}
\end{subequations}
\subsubsection{Lower-bound of PEP over $\left(0,\tau\right)$}

To attain the lower bound of $P_\tau\left(\mathbf{s}\rightarrow\hat{\mathbf{s}}\right)$, we need the following preparations.
We observe that $\mathbf{e}\mathbf{e}^T$ is rank-one with the only non-zero eigenvalue being $\lambda_{\max}=\sum_{j=1}^N e_j^2$.  Then,
 $\mathbf{h}_i^T\mathbf{e}\mathbf{e}^T\mathbf{h}_i\le \lambda_{\max}\sum_{j=1}^{N}h_{ij}^2,i=1,\ldots,M$. In other words, for all $\tau>0$, it holds that
\begin{subequations}
 \begin{eqnarray}
\{\mathbf{H}:\mathbf{h}_i\in\mathbb{R}^N_+,\lambda_{\max}\sum_{j=1}^{N}h_{ij}^2\le\tau^2,i=1,\ldots,M\}&&\nonumber\\
\subseteq\{\mathbf{H}:\mathbf{h}_i\in\mathbb{R}^N_+,\mathbf{h}_i^T\mathbf{e}\mathbf{e}^T\mathbf{h}_i\le \tau^2,i=1,\ldots,M\}&&
 \end{eqnarray}
  Further, it is true that
  \begin{eqnarray}\label{eqn:inner_super_rectangular}
  &&\{\mathbf{H}:0\le h_{ij}\le \frac{\tau}{\sqrt{N\lambda_{\max}}},j=1,\ldots,N,i=1,\ldots,M\}
  \nonumber\\
  &&\subseteq\{\mathbf{H}:\lambda_{\max}\sum_{j=1}^{N}h_{ij}^2\le\tau^2,i=1,\ldots,M\}
   \nonumber\\
  &&\subseteq\{\mathbf{H}:\mathbf{h}_i\in\mathbb{R}^N_+,\mathbf{h}_i^T\mathbf{e}\mathbf{e}^T\mathbf{h}_i\le \tau^2,i=1,\ldots,M\}
  \end{eqnarray}
  \end{subequations}
Now, by \eqref{eqn:inner_super_rectangular}, we can  lower-bound $P_{\tau}\left(\mathbf{s}\rightarrow\hat{\mathbf{s}}\right)$ by
\begin{subequations}
\begin{eqnarray}
&&P_{\tau}\left(\mathbf{s}\rightarrow\hat{\mathbf{s}}\right)\ge
\int_{h_{ij}\le\frac{\tau}{\sqrt{N\sum_{k=1}^Ne_k^2}}}P\left(\mathbf{s}\rightarrow\hat{\mathbf{s}}|\mathbf{H}\right)f_{\mathbf{H}}\left(\mathbf{H}\right)d\mathbf{H} \nonumber \\
&&\ge Q\left(\frac{\sqrt{M\rho}\tau}{2NP_{op}\sqrt{\sum_{k=1}^Ne_k^2}}\right)\int_{h_{ij}\le\frac{\tau}{\sqrt{N\sum_{k=1}^Ne_k^2}}}f_{\mathbf{H}}\left(\mathbf{H}\right)d\mathbf{H}\nonumber
\nonumber \\
&&\ge Q\left(\frac{\sqrt{M\rho}\tau}{2NP_{op}\ln \rho\sqrt{\sum_{k=1}^Ne_k^2}}\right)
\prod_{j=1}^N\prod_{i=1}^MQ\left(-\frac{\ln\frac{\tau}{\ln\rho\sqrt{N\sum_{k=1}^Ne_k^2}}}{\sigma_{ij}^2}\right)
\end{eqnarray}
where the last inequality holds for high SNR such that $\ln\rho>1$.
Again, by \eqref{eqn:q_function_lower_bound}, we arrive at the following lower-bound by
\begin{eqnarray}\label{eqn:lower_tau}
&&P_{\tau}\left(\mathbf{s}\rightarrow\hat{\mathbf{s}}\right)\ge \frac{\exp\left(\frac{MN}{2}\right)}{\left(4\pi\right)^{MN}}Q\left(\frac{\sqrt{M\rho}\tau}{2NP_{op}\ln\rho\sqrt{\sum_{k=1}^Ne_k^2}}\right)
\nonumber \\
&&\times\frac{\prod_{j=1}^N\prod_{i=1}^M\sigma_{ij}}{\ln^{MN}\frac{\ln\rho\sqrt{N\sum_{k=1}^Ne_k^2}}{\tau}}
\exp\left(-\sum_{j=1}^N\sum_{i=1}^M\frac{1}{2\sigma_{ij}^2}\left(\ln\frac{\tau}{\ln\rho\sqrt{N\sum_{k=1}^Ne_k^2}}\right)^2\right)
\end{eqnarray}
\end{subequations}

\subsubsection{Upper-bound of PEP over $\left(\tau,\infty\right)$}

Now, we are in a position to analyze $P\left(\mathbf{s}\rightarrow\hat{\mathbf{s}}\right)-P_{\tau}\left(\mathbf{s}\rightarrow\hat{\mathbf{s}}\right)$, i.e., the second term of \eqref{eqn:ml_detection_pep_fragment}, which is denoted by $\bar{P}_\tau\left(\mathbf{s}\rightarrow\hat{\mathbf{s}}\right)$.

  Similarly, for $h_{ij}$, $f'_{H}\left(h_{ij}\right)=0$ gives the extreme point $h_{ij,0}$  of $f_{H}\left(h_{ij}\right)$, i.e., $f_{H}\left(h_{ij}\right)\le f_{H}\left(h_{ij,0}\right),i=1,\ldots, M,j=1,\ldots,N$ . Then, we have $f_{\mathbf{H}}\left(\mathbf{H}\right)\le f_{\mathbf{H}}\left(\mathbf{H}_0\right)$ and thus, $\bar{P}_\tau\left(\mathbf{s}\rightarrow\hat{\mathbf{s}}\right)$ can be upper-bounded  by
  \begin{subequations}
\begin{eqnarray}
\bar{P}_\tau\left(\mathbf{s}\rightarrow\hat{\mathbf{s}}\right)\le \frac{f_{\mathbf{H}}\left(\mathbf{H}_0\right)}{2}
\int_{\left(\mathbf{e}^T\mathbf{h}_i\right)^2>\tau^2}
\exp\left(-\frac{\rho\sum_{i=1}^M\mathbf{h}_i^T\mathbf{e}\mathbf{e}^T\mathbf{h}_i}
{8NP_{op}^2}\right)d\mathbf{H}&
\end{eqnarray}

In fact, \eqref{eqn:inner_super_rectangular} also implies $\{\mathbf{H}:\mathbf{h}_i^T\mathbf{e}\mathbf{e}^T\mathbf{h}_i\ge \tau^2,i=1,\ldots,M\}\subseteq\{\mathbf{H}:h_{ij}\ge \frac{\tau}{\sqrt{N\sum_{k=1}^N e_k^2}},i=1,\ldots,M,j=1,\ldots,N\}$.
With this, $\bar{P}_\tau\left(\mathbf{s}\rightarrow\hat{\mathbf{s}}\right)$ can be further upper-bounded by
\begin{eqnarray}\label{eqn:upper_bound_tau_to_infty2}
&&\bar{P}_\tau\left(\mathbf{s}\rightarrow\hat{\mathbf{s}}\right)\le \frac{f_{\mathbf{H}}\left(\mathbf{H}_0\right)}{2}
\int_{h_{ij}\ge\frac{\tau}{\sqrt{N\sum_{k=1}^N e_k^2}}}
\exp\left(-\frac{\rho\lambda_{\max}\sum_{i=1}^M\sum_{j=1}^{N}h_{ij}^2}
{8NP_{op}^2}\right)d\mathbf{H}  \nonumber\\
&&=\frac{f_{\mathbf{H}}\left(\mathbf{H}_0\right)}{2}
\prod_{i=1}^M\prod_{j=1}^M\sqrt{2\pi}
\left(\frac{\rho\sum_{k=1}^Ne_k^2}{4NP_{op}^2}\right)^{-\frac{1}{2}}Q\left(\sqrt{\frac{\rho\tau^2}{4N^2P_{op}^2}}\right) \nonumber\\
 &&\le\frac{\exp\left(\frac{\sum_{i=1}^M\sum_{j=1}^N\sigma_{ij}^2}{2}\right)}{2\prod_{i=1}^M\prod_{j=1}^N\sigma_{ij}}
\left(\frac{\rho\sum_{k=1}^Ne_k^2}{NP_{op}^2}\right)^{-\frac{MN}{2}} \exp\left(-\frac{M\rho\tau^2}{8NP_{op}^2}\right)
\end{eqnarray}
\end{subequations}

So far, we have attained three bounds and now, examine their tightness to select $\tau$ in the following subsection.
\subsubsection{Selection of $\tau$}\label{subsec:tau}
It is noticed that the exponential terms of~\eqref{eqn:pep_upper_bound} and~\eqref{eqn:lower_tau} are the same, i.e., $\left(\ln \tau\right)^2\sum_{i=1}^M\sum_{j=1}^{N}\sigma_{ij}^{-2}$. To mathch the bounds~\eqref{eqn:pep_upper_bound} and  \eqref{eqn:upper_bound_tau_to_infty2}, $\tau$ is selected such that
\begin{eqnarray}\label{eqn:hyper_equation2}
\frac{M\rho\tau^2}{8NP_{op}^2}=\left(\ln \tau\right)^2\sum_{i=1}^M\sum_{j=1}^{N}\frac{\sigma_{ij}^{-2}}{2}
\end{eqnarray}
 Since a closed-form solution to~\eqref{eqn:hyper_equation2}  is hard to attain.
the solution to \eqref{eqn:hyper_equation2} is approximated below
  \begin{eqnarray}\label{eqn:tau_selection}
\tau=\sqrt{\frac{NP_{op}^2}{M}\sum_{i=1}^M\sum_{j=1}^{N} \sigma_{ij}^{-2}} \frac{\ln\rho}{\sqrt{\rho}}
 \end{eqnarray}
For the selection in~\eqref{eqn:tau_selection}, the following asymptotical equality holds.
 \begin{eqnarray*}
&&\lim_{\rho\rightarrow\infty}\frac{\frac{M\rho\tau^2}{8NP_{op}^2}}{\left(\ln \tau\right)^2\sum_{i=1}^M\sum_{j=1}^{N}\frac{\sigma_{ij}^{-2}}{2}}\nonumber\\
&&=\lim_{\rho\rightarrow\infty}\frac{\left(\ln \rho\right)^2}{\left(\ln \frac{\rho}{\ln^2 \rho}-\ln\left(\frac{NP_{op}^2}{M}\sum_{i=1}^M\sum_{j=1}^{N}\sigma_{ij}^{-2}\right)\right)^2}\nonumber\\
&&=1
 \end{eqnarray*}
Then, putting~\eqref{eqn:pep_upper_bound}, \eqref{eqn:lower_tau}, \eqref{eqn:upper_bound_tau_to_infty2} and \eqref{eqn:tau_selection} together,  we are allowed to arrive at the fact that there exists three positive constants $C_{2L}$, $C_{2U}$ and $C_{3U}$,  independent of $\rho$ shown by \eqref{eqn:pep_mimo}, postponed to the top of the next page,
  \begin{figure*}
  \begin{eqnarray}\label{eqn:pep_mimo}
  &&\underbrace{C_{2L} \left(\ln\rho\right)^{-MN}e^{-\sum_{i=1}^{M}\sum_{j=1}^{N}\frac{\left(\ln\rho +\ln \left(P_{op}^2\Omega\right)-\ln\left(M\sum_{k=1}^Ne_k^2\right)\right)^2}{8\sigma_{ij}^2}}}_{P_{L}\left(\mathbf{s}\rightarrow\hat{\mathbf{s}}\right)}
\le P\left(\mathbf{s}\rightarrow\hat{\mathbf{s}}\right)\nonumber\\
    &&\le\underbrace{ C_{2U}\left(\ln\rho\right)^{-MN}e^{-\sum_{i=1}^{M}\sum_{j=1}^{N}\frac{\left(\ln \frac{\rho}{\ln^2 \rho} +\ln \left(P_{op}^2\Omega\right)-\ln e_j^2\right)^2}{8\sigma_{ij}^2}}}_{P_{U1}\left(\mathbf{s}\rightarrow\hat{\mathbf{s}}\right)}+\underbrace{C_{3U}
\rho^{-\frac{MN}{2}}
e^{-\sum_{i=1}^{M}\sum_{j=1}^{N}\frac{\ln^2 \rho}{8\sigma_{ij}^2  }}}_{P_{U2}\left(\mathbf{s}\rightarrow\hat{\mathbf{s}}\right)}
\end{eqnarray}
    \hrule
\end{figure*}
where
\begin{subequations}
 \begin{eqnarray}
  \Omega=\sum_{i=1}^{M}\sum_{j=1}^{N}\sigma_{ij}^{-2}
  \end{eqnarray}
  \begin{eqnarray}
  C_{2L}=\frac{\prod_{i=1}^M\prod_{j=1}^N\sigma_{ij}}{\left(4\pi\right)^{MN}\exp\left(-\frac{MN}{2}\right)}Q\left(\frac{1}{2}\left(\sum_{k=1}^Ne_k^2\right)^{-\frac{1}{2}}\right)&& \end{eqnarray}
  \begin {eqnarray}
C_{2U}=\frac{\left(NP_{op}^2\right)^{MN}}{2\prod_{i=1}^M\prod_{j=1}^N\sqrt{\sigma_{ij}^2}}\exp\left(-\frac{\Omega}{8}\ln^2\left(\frac{NP_{op}^2\Omega}{M}\right)\right)
\end{eqnarray}
  \begin{eqnarray}
C_{3U}=\frac{\exp\left(\frac{\sum_{i=1}^M\sum_{j=1}^N\sigma_{ij}^2}{2}\right)}{2\prod_{i=1}^N\prod_{j=1}^M\sigma_{ij}}
\left(\frac{\sum_{k=1}^{N}e_k^2}{NP_{op}^2}\right)^{-\frac{MN}{2}}
 \end{eqnarray}
\end{subequations}
Now, we can see that in \eqref{eqn:pep_mimo}, $P_{L}\left(\mathbf{s}\rightarrow\hat{\mathbf{s}}\right)$ and $P_{U2}\left(\mathbf{s}\rightarrow\hat{\mathbf{s}}\right)$ have the same exponential term, $\exp\left(-\frac{\Omega}{8}\ln^2 \rho\right)$, whereas the exponential term of $P_{U1}\left(\mathbf{s}\rightarrow\hat{\mathbf{s}}\right)$ is $\exp\left(-\frac{\Omega}{8}\ln^2\frac{\rho}{\ln^2\rho}\right)$, which decays slower than  $\exp\left(-\frac{\Omega}{8}\ln^2 \rho\right)$ against high SNRs. That being said, we have attained the dominant term, $P_{U1}\left(\mathbf{s}\rightarrow\hat{\mathbf{s}}\right)$, of the upper-bound of $P\left(\mathbf{s}\rightarrow\hat{\mathbf{s}}\right)$. This completes the proof of Theorem~\ref{theorem:mimo_pep}.~\hfill$\Box$

\subsection{Proof of Property~\ref{property:local_two_worst_cases}}\label{app:local_last_two_worst_cases}
 For $m_1\perp n_1$, $m_2\perp n_2$ and $m_3\perp n_3$, we have that
 \begin{eqnarray*}
\frac{n_1+n_3}{m_1+m_3}\ge\frac{n_2}{m_2}\Leftrightarrow  m_2n_3-m_3n_2\ge m_1n_2-m_2n_1
 \end{eqnarray*}
 This is indeed true, since  we obtain $m_1n_2-m_2n_1=1$ by Lemma~\ref{lemma:farey_sequence} and $m_2n_3-m_3n_2\ge 1$ from the assumption that  $\frac{n_2}{m_2}<\frac{n_3}{m_3}$.

In the same token, given $m_0\perp n_0$, $m_1\perp n_1$ and $m_2\perp n_2$, we have that
\begin{eqnarray*}
\frac{n_0+n_2}{m_0+m_2}\le\frac{n_1}{m_1}
\Leftrightarrow m_0 n_1-m_1 n_0\ge m_1 n_2-m_2 n_1
\end{eqnarray*}

This indeed holds, since we can attain $m_1n_2-m_2n_1=1$  by Lemma~\ref{lemma:farey_sequence} and $m_0 n_1-m_1 n_0\ge 1$ from the assumption that  $\frac{n_0}{m_0}<\frac{n_1}{m_1}$. This completes the proof of Property~\ref{property:local_two_worst_cases}.
~\hfill $\Box$

\subsection{Proof of Property~\ref{property:the_local_solution}}\label{app:the_local_solution}
First, Statement~1: $\frac{n_1}{m_1}<\frac{n_1+n_2}{m_1+m_2}<\frac{n_2}{m_2}$ can be verified by
    \begin{eqnarray*}
     \frac{n_1+n_2}{m_1+m_2}-\frac{n_2}{m_2}
    =\frac{m_1}{\left(m_1+m_2\right)}\left(\frac{n_1}{m_1}-\frac{n_2}{m_2}\right)<0, \\
        \frac{n_1+n_2}{m_1+m_2}-\frac{n_1}{m_1}
  =\frac{m_2}{\left(m_1+m_2\right)}\left(\frac{n_2}{m_2}-\frac{n_1}{m_1}\right)>0.
     \end{eqnarray*}
     When $\frac{f_{11}}{f_{12}},\frac{f_{21}}{f_{22}}\in\left(\frac{n_1}{m_1},\frac{n_2}{m_2} \right)$,
     we can have $ \left(m_1f_{11}-n_1f_{12}\right)>0$ and
      $ \left(m_2f_{11}-n_2f_{12}\right)<0$.
Hence,
\begin{subequations}
        \begin{eqnarray}\label{eqn:former_part}
      | m_1f_{11}-n_1f_{12}|-|m_2f_{11}-n_2f_{12}|
    =\left(m_1+m_2\right)f_{12}\left(\frac{f_{11}} {f_{12}}-\frac{n_1+n_2}{m_1+m_2}\right)
     \end{eqnarray}
   and
             \begin{eqnarray}\label{eqn:latter_part}
     | m_1f_{21}-n_1f_{22}|-|m_2f_{21}-n_2f_{22}|=\left(m_1+m_2\right)f_{22}\left(\frac{f_{21}} {f_{22}}-\frac{n_1+n_2}{m_1+m_2}\right).
     \end{eqnarray}
     \end{subequations}
Combining \eqref{eqn:former_part} and \eqref{eqn:latter_part} leads us to the following facts:
\begin{enumerate}
      \item If $\frac{f_{11}}{f_{12}},\frac{f_{21}}{f_{22}}\in\left(\frac{n_1}{m_1},\frac{n_1+n_2}{m_1+m_2}\right)$, then, we have
      \begin{eqnarray*}
-m_2f_{11}+n_2f_{12}>m_1f_{11}-n_1f_{12}>0,-m_2f_{21}+n_2f_{22}>m_1f_{21}-n_1f_{22}>0
          \end{eqnarray*}

      Thus, $F_{m_1n_1}<F_{m_2n_2}$ holds.
      \item If $\frac{f_{11}}{f_{12}},\frac{f_{21}}{f_{22}}\in\left(\frac{n_1+n_2}{m_1+m_2},\frac{n_2}{m_2}\right)$, then, we arrive at
           \begin{eqnarray*}
            &&m_1f_{11}-n_1f_{12}>-m_2f_{11}+n_2f_{12}>0\\
            &&m_1f_{21}-n_1f_{22}>-m_2f_{21}+n_2f_{22}>0.
          \end{eqnarray*}
           Hence, we have $F_{m_1n_1}>F_{m_2n_2}$.
      \item If $\frac{f_{11}}{f_{12}}=\frac{f_{21}}{f_{22}}=\frac{n_1+n_2}{m_1+m_2}$, then $F_{m_1n_1}=F_{m_2n_2}$.
\end{enumerate}
    This completes the proof of Property~\ref{property:the_local_solution}.
~\hfill $\Box$

\subsection{proof of Property~\ref{th:min}}\label{app:th:min}
$\forall \frac{n_0}{m_0},\frac {n_3}{ m_3}\in \mathfrak{F}_{k}$ satisfying $\frac{n_0}{m_0}<\frac {n_1}{ m_1}<\frac{n_2}{ m_2}<\frac{n_3}{ m_3}$,
when $\frac{f_{11}}{f_{12}},\frac{f_{21}}{f_{22}}\in\left(\frac{n_1}{m_1},\frac{n_2}{m_2} \right)$, we have
$F_{m_0n_0},F_{m_1n_1},F_{m_2n_2},F_{m_3n_3}>0$.
Property \ref{property:local_two_worst_cases}
tells us that  $\frac{n_1+n_3}{m_1+m_3}\ge\frac{n_2}{m_2}$ and $\frac{n_0+n_2}{m_0+m_2}\le\frac{n_1}{m_1}$.
As a result, we can have
 \begin{eqnarray*}
  &\left(\frac{n_1}{m_1},\frac{n_2}{m_2} \right)\subseteq \left(\frac{n_0+n_2}{m_0+m_2},\frac{n_2}{m_2} \right)\subseteq \left(\frac{n_0}{m_0},\frac{n_2}{m_2} \right)&\\
   &\left(\frac{n_1}{m_1},\frac{n_2}{m_2} \right)\subseteq \left(\frac{n_1}{m_1},\frac{n_1+n_3}{m_1+m_3} \right)\subseteq \left(\frac{n_1}{m_1},\frac{n_3}{m_3} \right)&
 \end{eqnarray*}

Using Property \ref{property:the_local_solution}, we can attain $ F_{m_0n_0}>F_{m_2n_2}$ and  $ F_{m_1n_1}<F_{m_3n_3}$. The successivity of $\frac{n_1}{m_1}$ and $\frac{n_2}{m_2}$ ensures $F_{m_1n_1}$ and $F_{m_2n_2}$ are  the last two minimum over $\left(\frac{n_1}{m_1},\frac{n_2}{m_2} \right)$. This completes the proof of Property~\ref{th:min}.
~\hfill $\Box$

\subsection{proof of Theorem~\ref{theorem:golbal_solution}}\label{app:global_solution}
Assume $\frac{n_1}{m_1}$ and $\frac{n_2}{ m_2}$  are varying two successive items of $\mathfrak{F}_{2^p-1}$ and denote $x=\frac{f_{11}}{f_{12}}$ and $y=\frac{f_{21}}{f_{22}}$.
These notations produce $f_{12}\left(1+x\right)+f_{22}\left(1+y\right)=1$. We firstly prove that if $x,y\in \left(\frac{n_1}{m_1},\frac{n_2}{m_2} \right)$, then the local solution  to~\eqref{eqn:modulator_design} is determined by
\begin{eqnarray}\label{theorem:local_solution}
x&=&y=\frac{n_1+n_2}{m_1+m_2},\nonumber \\
f_{12}&=&f_{22}=\frac{m_1+m_2}{2\left(m_1+m_2+n_1+n_2\right)}
\end{eqnarray}
 Letting $\left(xm_1-n_1\right)\left(ym_1-n_1\right)\le  \left(xm_2-n_2\right)\left(ym_2-n_2\right)
 $ yields $x\le \frac{n_1\left(ym_1-n_1\right)-n_2\left(ym_2-n_2\right)}{m_1\left(ym_1-n_1\right)-m_2\left(ym_2-n_2\right)}
 $. In the same token, $\left(xm_1-n_1\right)\left(ym_1-n_1\right)\ge  \left(xm_2-n_2\right)\left(ym_2-n_2\right)
 $ implies $y\le \frac{n_2\left(xm_2-n_2\right)-n_1\left(xm_1-n_1\right)}{m_2\left(xm_2-n_2\right)-m_1\left(xm_1-n_1\right)}
 $.
  Then,
from Property~\ref{property:the_local_solution} and Property~\ref{th:min}, we have that if $x,y\in \left(\frac{n_1}{m_1},\frac{n_2}{ m_2}\right)$, then \eqref{eqn:modulator_design} can be equivalent to the following two sub-problems:
\begin{enumerate}
  \item If $x\le \frac{n_1\left(ym_1-n_1\right)-n_2\left(ym_2-n_2\right)}{m_1\left(ym_1-n_1\right)-m_2\left(ym_2-n_2\right)}
 $, then  \eqref{eqn:modulator_design} is equivalent to the following problem
  \begin{eqnarray}\label{eqn:modulator_design_sub1}
&&\max_{f_{11},f_{12},f_{21},f_{22}} F_{m_1n_1}\nonumber \\
&& s.t.
\left\{
  \begin{array}{ll}
  x\le \frac{n_1\left(ym_1-n_1\right)-n_2\left(ym_2-n_2\right)}{m_1\left(ym_1-n_1\right)-m_2\left(ym_2-n_2\right)}
,\\
x,y\in \left(\frac{n_1}{m_1},\frac{n_2}{m_2} \right),\\
f_{11}+f_{12}+f_{21}+f_{22}=1.
  \end{array}
\right.
\end{eqnarray}
Over the feasible region of $x$ and with the increasing property of $F_{m_1n_1}$, we can have
\begin{eqnarray*}
&&F_{m_1n_1}\le \left(m_1\frac{n_1\left(ym_1-n_1\right)-n_2\left(ym_2-n_2\right)}{m_1\left(ym_1-n_1\right)-m_2\left(ym_2-n_2\right)}-n_1\right) f_{12}f_{22}\left(ym_1-n_1\right)
 \nonumber \\
 &&=\frac{m_2y\left(m_2n_1-m_1n_2\right)+n_2\left(m_1n_2-m_2n_1\right)}{{m_1\left(ym_1-n_1\right)-m_2\left(ym_2-n_2\right)}}
f_{12}f_{22}\left(ym_1-n_1\right)
 \nonumber \\
 &&=\frac{-f_{12}f_{22}\left(ym_2-n_2\right)\left(ym_1-n_1\right)}{{m_1\left(ym_1-n_1\right)-m_2\left(ym_2-n_2\right)}}
\end{eqnarray*}
where the last equality is attained by using $m_1n_2-m_2n_1=1$ and the equality holds when $x= \frac{n_1\left(ym_1-n_1\right)-n_2\left(ym_2-n_2\right)}{m_1\left(ym_1-n_1\right)-m_2\left(ym_2-n_2\right)}$ .
This upper bound of $\frac{F_{m_1n_1,\mathbf{F}}}{f_{12}f_{22}}$ is denoted by $f\left(y\right)$ and its derivative is
\begin{eqnarray*}
f'\left(y\right)
=-\frac{\left(\left(m_1+m_2\right)y-\left(n_1+n_2\right)\right)}{\left({m_1\left(ym_1-n_1\right)-m_2\left(ym_2-n_2\right)}\right)^2}\left(\left(m_1-m_2\right)y-\left(n_1-n_2\right)\right)
\end{eqnarray*}
Letting $f'\left(y\right)=0$
 yields that $y=\frac{n_1-n_2}{m_1-m_2}$ or $\frac{n_1+n_2}{m_1+m_2}$.
 Now, we prove $y=\frac{n_1+n_2}{m_1+m_2}$ is the maximum point. Firstly, we verify that $\frac{n_1-n_2}{m_1-m_2}$ is outside $\left(\frac{n_1}{m_1},\frac{n_2}{m_2} \right)$. We can have
 \begin{eqnarray*}
  \frac{n_1-n_2}{m_1-m_2}-\frac{n_1}{m_1}=\frac{-1}{m_1\left(m_1-m_2\right)}\\
  \frac{n_1-n_2}{m_1-m_2}-\frac{n_2}{m_2}=\frac{-1}{m_2\left(m_1-m_2\right)}
 \end{eqnarray*}
 From \cite{hardy1979introduction}, we know $m_1\neq m_2$. Therefore, for the sign of $m_1-m_2$, we have the following two possibilities
 \begin{enumerate}
   \item If $m_1>m_2$, then $\frac{n_1-n_2}{m_1-m_2}<\frac{n_1}{m_1}<\frac{n_2}{m_2}$. We can have that over $\left(\frac{n_1}{m_1},\frac{n_1+n_2}{m_1+m_2}\right)$, $f'\left(y\right)$ is positive and over $\left(\frac{n_1+n_2}{m_1+m_2},\frac{n_2}{m_2}\right)$, $f'\left(y\right)$ is negative.
   \item If $m_1<m_2$, then $\frac{n_1-n_2}{m_1-m_2}>\frac{n_2}{m_2}>\frac{n_1}{m_1}$. Also, we can have that over $\left(\frac{n_1}{m_1},\frac{n_1+n_2}{m_1+m_2}\right)$, $f'\left(y\right)$ is positive and over $\left(\frac{n_1+n_2}{m_1+m_2},\frac{n_2}{m_2}\right)$, $f'\left(y\right)$ is negative.
 \end{enumerate}
 These observations indicate that $y=\frac{n_1+n_2}{m_1+m_2}$ is the maximum point. Combining  $y=\frac{n_1+n_2}{m_1+m_2}$ and $x= \frac{n_1\left(ym_1-n_1\right)-n_2\left(ym_2-n_2\right)}{m_1\left(ym_1-n_1\right)-m_2\left(ym_2-n_2\right)}$ gives $x=y=\frac{n_1+n_2}{m_1+m_2}$.
  \item If $y\le \frac{n_2\left(xm_2-n_2\right)-n_1\left(xm_1-n_1\right)}{m_2\left(xm_2-n_2\right)-m_1\left(xm_1-n_1\right)}
 $, then  \eqref{eqn:modulator_design} is transformed into the following problem
  \begin{eqnarray}\label{eqn:modulator_design_sub2}
&&\max_{f_{11},f_{12},f_{21},f_{22}} F_{m_2n_2}\nonumber \\
&& s.t.
\left\{
  \begin{array}{ll}
  y\le \frac{n_2\left(xm_2-n_2\right)-n_1\left(xm_1-n_1\right)}{m_2\left(xm_2-n_2\right)-m_1\left(xm_1-n_1\right)},\\
x,y\in \left(\frac{n_1}{m_1},\frac{n_2}{m_2} \right),\\
f_{11}+f_{12}+f_{21}+f_{22}=1.
  \end{array}
\right.
\end{eqnarray}
 In the same token, we can have the solution to \eqref{eqn:modulator_design_sub2} is
 $x=y=\frac{n_1+n_2}{m_1+m_2}$.
\end{enumerate}

Putting things together tells us that with $x,y\in \left(\frac{n_1}{m_1},\frac{n_2}{ m_2}\right)$ the local solution to \eqref{eqn:modulator_design} is
\begin{eqnarray}\label{eqn:local_solution}
\frac{f_{11}}{f_{12}}=\frac{f_{21}}{f_{22}}=\frac{n_1+n_2}{m_1+m_2}
\end{eqnarray}
It follows that $F_{m_1n_1}=F_{m_2n_2}$ from Property \ref{property:the_local_solution}.

    Now, it is time to give the analytical FDSC.
Combining \eqref{eqn:local_solution} with $f_{11}+f_{12}+f_{21}+f_{22}=1$ produces $f_{12}+f_{22}=\frac{m_1+m_2}{m_1+m_2+n_1+n_2}$. In addition, using the geometrical and arithmetical inequality: $\sqrt{ab}\le \frac{a+b}{2}$ for $a, b\ge 0$, we obtain
 \begin{eqnarray*}
F_{m_1n_1}&=&f_{12}f_{22}\left(m_1x-n_1\right)\left(m_1y-n_1\right)\nonumber \\
&\le& \frac{\left(f_{12}+f_{22}\right)^2}{4}\left(\frac{m_1n_2-m_2n_1}{m_1+m_2}\right)^2\nonumber \\
 &=&\frac{\left(m_1n_2-m_2n_1\right)^2}{4\left(m_1+m_2+n_1+n_2\right)^2}
 \end{eqnarray*}
 where the equality holds if and only if $f_{11}=f_{21}=\frac{n_1+n_2}{2\left(m_1+m_2+n_1+n_2\right)}$ and $f_{12}=f_{22}=\frac{m_1+m_2}{2\left(m_1+m_2+n_1+n_2\right)}$. Then, the local maximum value of the objective function is given by
\begin{eqnarray}\label{eqn:local_value}
 \max_{x,y\in \left(\frac{n_1}{m_1},\frac{n_2}{m_2} \right)}\min_{\mathbf{e}} e_1e_2 =\frac{1}{4\left(m_1+m_2+n_1+n_2\right)^2}.
\end{eqnarray}
where we use $m_1n_2-m_2 n_1=1$ from Proposition~\ref{lemma:farey_sequence}.
On the other hand,   Proposition~\ref{lemma:farey_sequence} reveals $
 m_1+m_2+n_1+n_2\ge 1+2^{p}$.
 This inequality implies
  \begin{eqnarray} \label{eqn:global_maximum}
  \max_{\mathbf{F}}\min_{\mathbf{e}} e_1e_2  \le \frac{1}{4\left(1+2^{p}\right)^2}.
  \end{eqnarray}
where the equality holds when $\frac{f_{11}}{f_{12}}=\frac{f_{21}}{f_{22}}=\frac{1}{2^p}$
or $\frac{f_{11}}{f_{12}}=\frac{f_{21}}{f_{22}}=2^p$. Further, using $\sum_{i,j=1}^2 f_{ij}=1$ gives  $f_{11}=f_{21}=\frac{2^{p-1}}{2^p+1}$.
 Finally, the global  solution  to \eqref{eqn:modulator_design} is given by
\begin{eqnarray*}
  \mathbf{F}_{FDSC} =\frac{1}{2+2^{p+1}}\left(
  {\begin{array}{ccc}
  2^p&1\\
  2^p&1\\
 \end{array}}
  \right).
\end{eqnarray*}
 Until now, our  focus is on the sequence  before $1/ 1$. Using   the reciprocity  of the breakpoint sequence gives the final solution shown by \eqref{eqn:global_optimal_modulator}.
This completes the proof of Theorem~\ref{theorem:golbal_solution}.
~\hfill $\Box$

\bibliographystyle{ieeetr}
\bibliography{tzzt}
\end{document}